%% file: Lundqvist_pwn0540_corrected.tex
\documentclass[useAMS,usenatbib]{mn2e}
\usepackage{txfonts,amssymb}
\usepackage{graphicx}
\usepackage{hyperref}
\usepackage{afterpage,array,longtable}

\def\asec{\ifmmode ^{\prime\prime}\else$^{\prime\prime}$\fi}
\def\etal{{et\,al. }}
\def\msun{M$_{\odot}$}

\def\grad{$^\circ$}
\def\degs{\ifmmode ^{\circ}\else$^{\circ}$\fi}
\def\amin{\ifmmode ^{\prime}\else$^{\prime}$\fi}
\def\asec{\ifmmode ^{\prime\prime}\else$^{\prime\prime}$\fi}
\def\farcs{\hbox{$.\!\!^{\prime\prime}$}}  

\def\degs{\ifmmode ^{\circ}\else$^{\circ}$\fi}
\def\amin{\ifmmode ^{\prime}\else$^{\prime}$\fi}

\def\aap{{A\&A\/}}

\def\cm3{\rm ~cm^{-3}}
\def\kms{\rm ~km~s^{-1}}

\def\wl{~\lambda}
\def\wll{~\lambda\lambda}

\def\Msun{~{\rm M}_\odot}
\def\Ti44{M(^{44}{\rm Ti})}

\def\psr{PSR~B0540-69.3}
\def\snr{SNR~0540-69.3}

\def\lsim{\!\!\!\phantom{\le}\smash{\buildrel{}\over
  {\lower2.5dd\hbox{$\buildrel{\lower2dd\hbox{$\displaystyle<$}}\over
                               \sim$}}}\,\,}
\def\gsim{\!\!\!\phantom{\ge}\smash{\buildrel{}\over
  {\lower2.5dd\hbox{$\buildrel{\lower2dd\hbox{$\displaystyle>$}}\over
                               \sim$}}}\,\,}

\title[Evolution and polarization of the pulsar wind nebula of PSR B0540-69.3]
{Spectral evolution and polarization of variable structures in the pulsar wind nebula of PSR B0540-69.3}

\author[N.~Lundqvist et al.]{N.~Lundqvist$^{1}$\thanks{E-mail:natalia@astro.su.se},
P.~Lundqvist$^{1}$, C.-I.~Bj\"ornsson$^{1}$, G.~Olofsson$^{1}$, S.~Pires$^{2}$, 
\newauthor
Yu.~A.~Shibanov$^{3,4}$, D.~A.~Zyuzin$^{3,5}$\\
$^{1}$Department of Astronomy, Stockholm University, AlbaNova Science 
Center, SE-106 91 Stockholm, Sweden\\
$^{2}$CEA/DSM/DAPNIA/SEDI, CE Saclay, 91191 Gif-sur-Yvette, France\\
$^{3}$Ioffe Physical Technical Institute, Politekhnicheskaya 26, 
St. Petersburg, 194021, Russia\\
$^{4}$St. Petersburg State Polytechnical Univ., Polytekhnicheskaya 29, St.
Petersburg, 195251, Russia\\
$^{5}$Academic University, Khlopina 8, St. Petersburg, 194021, Russia\\}

\date{Accepted 2010}

\pagerange{\pageref{firstpage}--\pageref{lastpage}} \pubyear{2010}

\begin{document}
\label{firstpage}
\maketitle

\begin{abstract}
We present high spatial resolution optical imaging and polarization observations of the
\psr\ and its highly dynamical pulsar wind nebula (PWN) performed with 
{\it Hubble Space Telescope}, and compare them with X-ray data obtained with the
{\it Chandra X-ray Observatory}.  In particular, we have studied the bright region southwest 
of the pulsar where a bright ``blob" is seen in 1999. In a recent paper by De Luca et al. it was
argued that the ``blob" moves away from the pulsar at high speed. We show that it
may instead be a result of local energy deposition around 1999, and that the emission
from this then faded away rather than moved outward. Polarization data from 2007
show that the polarization properties show dramatic spatial variations at the 1999 blob position
arguing for a local process. Several other positions along the pulsar-``blob" orientation
show similar changes in polarization, indicating previous recent local energy depositions.
In X-rays, the spectrum
steepens away from the ``blob" position, faster orthogonal to the pulsar-``blob" 
direction than along this axis of orientation. This could indicate that the pulsar-``blob" orientation
is an axis along where energy in the PWN is mainly injected, and that this is then mediated
to the filaments in the PWN by shocks. We highlight this by constructing an [S~II]-to-[O~III]-ratio
map, and comparing this to optical continuum and X-ray emission maps. We argue,
through modeling, that the high [S~II]/[O~III] ratio is not due to time-dependent photoionization
caused by possible rapid X-ray emission variations in the ``blob" region. We have also
created a multiwavelength energy spectrum for the ``blob" position showing that one can, to 
within 2$\sigma$, connect the optical and X-ray emission by a single power law. The slope of 
that power-law (defined from $F_{\nu} = \nu^{-\alpha_{\nu}}$) would 
be $\alpha_{\nu}=0.74\pm0.03$, which is marginally different from the X-ray spectral slope 
alone with $\alpha_{\nu}=0.65\pm0.03$.  A single power-law for most of the PWN is, 
however, not be possible. We obtain best power-law fits for the X-ray
spectrum if we include ``extra" oxygen, in addition to the oxygen column density in 
the interstellar gas of the Large Magellanic Cloud and the Milky Way. This oxygen is most
naturally explained by the oxygen-rich ejecta of the supernova remnant. The oxygen needed
likely places the progenitor mass in the $20-25 \Msun$ range, i.e., in the upper mass
range for progenitors of Type IIP supernovae.

\end{abstract}


\section{Introduction}

\psr\ is a 50.2 millisecond pulsar in the Large Magellanic Cloud (LMC). A 
detailed review about the discovery and observations of the pulsar, its wind
nebula, as well as the whole supernova remnant in all wavelength bands is 
provided by \citet{Williams08}. 

\psr\ is often referred to as the ``Crab twin'' because it is of similar age 
($\sim 1000$ years), it has a similar pulse period and that it has a pulsar wind nebula 
(PWN) surrounding it, similar to that of the Crab. It has also been suggested that the 
detailed structures of the two PWNe, as revealed for \psr\ in X-rays  \citep{GW00}, are
similar, and that \psr\ may have a jet and a torus like the Crab. 
For further comparison between the two pulsars and their PWNe we refer to
\citet{Seraf04}.

There are, however, important differences between the two pulsars and their
surrounding nebulae. The most obvious is that the supernova ejecta of
\snr\ are oxygen-rich, although with some hydrogen mixed in \citep{Seraf05}, 
making it likely that \snr\ and \psr\ (together henceforth referred to as ``0540'') stem from 
a Type IIP supernova explosion of a massive ($\sim 20$ \msun) star \citep{Chevalier06}, 
whereas the Crab progenitor was a much less massive (8-10 \msun) star \citep[e.g.,][]{Hester08}. 
In addition, 0540 has a normal supernova shell of fast ejecta, something which has 
not yet been fully confirmed to exist around the Crab \citep[see, e.g.,][]{Tziamtzis09}. 
This could mean that the Crab ejecta have much less kinetic energy than the ejecta
of 0540 which appears to be normal in that respect. 

Closer to the center, 0540 appears much more asymmetric than
the Crab Nebula, with much of the emission coming from a region a few arcseconds 
southwest of the pulsar. \citet{Morse06} show that this asymmetric appearance,
as well as red-ward asymmetry in integrated profiles of 
emission lines \citep[see also][]{Kirshner89,Seraf05}, 
does not signal an overall asymmetry of the PWN as the maximum velocity
of emission lines toward and away from us are nearly equal.

Even closer to the center, a detailed comparison between the PWNe of the
two pulsars is more difficult due to the large distance to \psr\ 
\citep[$\sim 51$ kpc,][]{Panagia05} as opposed to the $2\pm 0.5$ kpc to the 
Crab \citep{Kaplan08} and references there in); the semi-major
axis of the torus in the Crab would only subtend $\sim 1\farcs5$ at the distance
of \psr, and details such as the changing structure of the Crab wisps would have 
been close to impossible to disentangle even with the HST or ground-based adaptive
optics. It therefore comes as no surprise that there is yet no counterpart for
\psr\ and its PWN to the detailed synchronized optical and X-ray study of the 
Crab PWN by \citet{Hester02}. These authors revealed the presence of outward 
moving equatorial wisps with velocities of $\sim$~0.5 the speed of light. Many 
time-variable subarcsecond structures exist in the Crab PWN in the optical 
\citep{Hester02}, near-IR \citep{Melatos05} and X-rays \citep{Hester02,Weisskopf00}.
This is discussed in detail by \citet{Hester08}. 

The first indication that also 0540 displays changes in its structure over a relatively
short time period was shown by \citet{DeLuca07}. They found a 10 year flux variation
(between 1995 and 2005) in the southwest direction where the PWN has its strongest 
emission and suggested that they had detected a hot spot moving at $\sim 0.04c$.
De Luca et al. noted that the hot spot could be similar to a time-varying arc-like feature 
in the outer Crab Nebula, and that a pulsar jet in 0540 could be  directed toward the 
bright southwest region rather than perpendicular to this, as suggested earlier by 
\citet{GW00}.

In this paper we present a study of the PWN of \psr\ using the same data as 
\citet{DeLuca07} , complemented with all other HST data available for 0540 as well.
In particular, we have included recent HST polarization data from 2007. We also
include all available X-ray data of 0540. 

The paper is organized as follows: in Sect. 2 we describe and discuss the observations,
in Sect. 3 the analysis and results, and in Sect. 4 we conclude with a discussion.

\section{Observations}

\subsection{HST observations and data analysis}

The  pulsar field was obseved with HST/WFPC2 in 2005 (Program ID 10601) 
using the two broadband filters F547M and F555W (see Table~\ref{t21:arch}). 
As opposed to earlier HST/WFPC2 observations of \psr\ we used a dithering procedure 
for all images to better avoid cosmic ray contamination. \psr\ and its PWN 
were exposed on the PC chip, and the observations were made in a two-gyro mode. 
The CCD amplification, i.e., {\sf gain}  was 7.12 and {\sf readout noise}=5.24. 
The reason for choosing the continuum filter F547M was that the main goals 
of the observations were to study the proper motion of the pulsar.
as well as the spectrum of the pulsar and the PWN. 
In earlier observations the pulsar and PWN were mainly studied in filters 
largely contaminated with line emission from filaments in 0540. 
 Continuum filters like F547M avoid this contamination. The 2005 data have been 
discussed by De Luca et al. (2007), and an upper limit to the proper motion of the pulsar 
was estimated, as well as identifying the apparent displacement of a feature 
in the PWN. We discuss this further here, as well as use the data for 
spectral studies of the PWN.
The
\input{tab21.tex}

To study time variability of the PWN  we also used available  
archival HST/WFPC and WFPC2 data obtained from 1992 to 2007   
(Table~\ref{t21:arch}).  
There are three different storage sites of  HST observations, namely
the {\sf MAST}\footnotemark\footnotetext{\tiny Multimission Archive at 
STScI, \href{archive.stsci.edu}{archive.stsci.edu}}, 
{\sf CADC}\footnotemark\footnotetext{\tiny Hubble Space Telescope Archive 
at Canadian Astronomy Data Center, \href{www4.cadc-ccda.hia-iha.nrc-cnrc.gc.ca/hst}
{www4.cadc-ccda.hia-iha.nrc-cnrc.gc.ca/hst} }, 
and {\sf ST-ECF}\footnotemark\footnotetext{\tiny Space Telescope $–$ European 
Coordinating Facility, \href{archive.eso.org/cms/hubble-space-telescope-data}
{archive.eso.org/cms/hubble-space-telescope-data}}. 
All three archives use slightly different calibration plans, or pipe-line 
reductions of raw data. We found  that the  pipe-line reduced  data of the 
same WFPC2 observations from these archives give different results for the 
photometry of \psr\ and its neighborhood. In previous work 
(Serafimovich et. al. 2004) we did not resort to pipe-line reductions, but 
made careful manual reductions of the raw data. Using  photometric results 
from that work we found that only the data from the {\sf MAST} archive 
with their recent pipe-line calibration plan can reproduce our previous photometry.  
To avoid systematic differences due to pipe-line effects in our analysis we 
therefore used the data only from the {\sf MAST} archive with the most 
recent calibration plan for all epochs of HST/WFPC2 observations listed in 
Table~\ref{t21:arch}. 

\begin{figure}
\begin{center}
\includegraphics[width=80mm, clip]{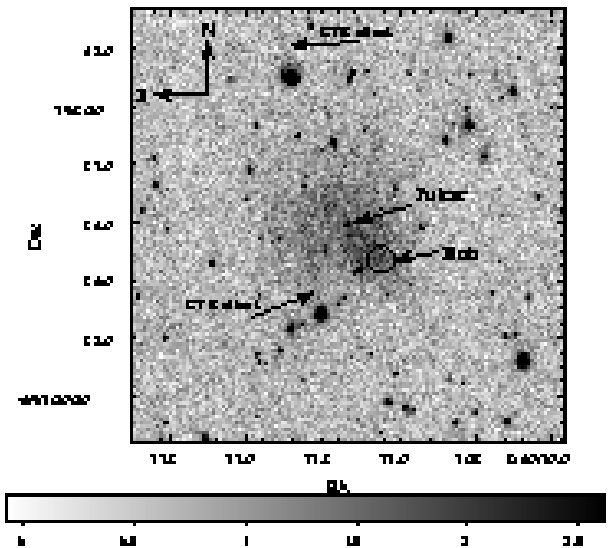}
\end{center}
\caption{A 15\arcsec$\times$15\arcsec\ image of the field
around \psr~obtained in the F547M band with HST/WFPC2 in 2005 (Table~\ref{t21:arch}).
The pulsar was exposed on the PC chip and its position is marked by an
arrow. The diffuse emission surrounding the pulsar is the pulsar wind nebula 
(PWN). A highly variable area or ``blob'' is marked by a circle and an arrow. 
Two typical traces of the CTE effect are marked by arrows (see text for more 
details).
} \label{f:F547im}
\end{figure}

All data were reduced using {\sf IRAF} in combination with {\sf STSDAS and 
DITHER}\footnotemark\footnotetext{\tiny{Space Telescope Science Data 
Analysis System, \href{www.stsci.edu/hst/wfpc2/documents/dither\_handbook.html}
{www.stsci.edu/hst/wfpc2/documents/dither\_handbook.html}}} packages. 
Pre-calibrated data include references to the latest distortion correction 
files available at the STScI\footnotemark\footnotetext{\tiny{\href{ftp.stsci.
edu/cdbs/uref/}{ftp.stsci.edu/cdbs/uref/}}}. Using STScI 
recommendations\footnotemark\footnotetext{\tiny{\href{incubator.stsci.
edu/mediawiki/index.php/WFPC2}{incubator.stsci.edu/mediawiki/index.php/WFPC2}}} 
we used these files together with the {\sf Multidrizzle} package for each 
individual image independently. Cosmic ray cleaning and combining of 
the images were excluded at this step. This exclusion was essential for the 1995 
and 1999 epochs. These observations were performed without dithering in 
between exposures, and after cleaning and combining, {\sf Multidrizzle} 
introduced extra noise in the pixel distribution which led to an overestimate 
of faint source fluxes. The overestimate for F547M in the 1999 epoch was 
found to be a factor of around two. After {\sf Multidrizzle}, the individual images 
were converted from the unit of electrons per second to counts using the
{\sf IRAF/imgtools/imcalc} utility, e.g., $(image*exposure)/gain$. Final 
image combining was performed using the standard {\sf IRAF imcombine} utility 
with cosmic ray rejection by setting {\sf reject}$=${\sf crreject}.

The data obtained in 2005, and later, clearly show that the Charge Transfer 
Efficiency (CTE) effect (\cite{riess00}) on HST/WFPC2 images was getting 
worse with time (see Figure~\ref{f:F547im}). The CTE effect is very
complicated to compensate for as it is time, position and flux dependent, i.e.,
it is unique for every particular point source and almost unknown for extended 
objects. In order to more reliably study the structure of the PWN, we used the 
multi-scale filtering method described in Sect.~\ref{Wavel} and~\ref{CTEpm}. 

\subsection{Chandra X-ray observations}\label{chandra}
\input{tab22.tex}
To compare the structure of the PWN in the optical and X-rays, we utilized 
the Chandra archival data listed in Table~\ref{t3:arch}. 
 The HRC-I observations of 0540 were obtained as part of the instrument 
calibration plan in 1999 and 2000, only a few months before and after the 
HST 1999 observations described in Table~\ref{t21:arch}. They  contain very 
limited spectral information, but provide the best spatial resolution 
(pixel size 0\farcs132) in X-rays, and are therefore the most useful for the study 
of the PWN  morphology. The ACIS-S observations (PI S. Park) have lower 
spatial resolution (pixel size 0\farcs492), but provide better spectral information.   
The ACIS-S observations started only three months after our HST observations in 
2005, and are useful for an optical/X-ray comparison, assuming the  PWN structure 
did not change substantially in three months.  Another advantage  
of these observations is that the 1/4 ACIS-S subarray mode was  
used\footnotemark\footnotetext{\tiny{The frame time of the 1/4 mode is 
$\la$1.1 s, for details see \href{asc.harvard.edu/proposer/POG/html/chap6.html}
{asc.harvard.edu/proposer/POG/html/chap6.html}}} and the data are much less 
contaminated by the CCD pileup effects from the high count rate of the pulsar 
in the center of the PWN, as opposed to previous ACIS observations, e.g., 
the data obtained in Nov 23 1999 (\cite{Petre07},  Obs ID 119, exposure 28.16 ks).
For the latter, significant pileup precludes a reliable spatial and spectral analysis  
of the central PWN regions. The HRC-I data are not affected by the pileup.    

\subsection{HST/WFPC2 polarization observations}\label{polariz}

To better understand the highly variable structure in the
southwestern part of the 0540 PWN we utilized HST/WFPC2 archival 
polarization data. 

\input{tab23.tex}

\psr\ and its wind nebula were observed in polarized light with
HST/WFPC2 in 2007 (Program ID 10900) using the broad band filter F606W in 
combination with the POLQ polarizer (see Table~\ref{t23:arch}). The field was 
exposed only on the PC chip during all sets of observations using different
rotations of the HST to reproduce the required polarization angle (see column 4 in
Table~\ref{t23:arch}). This gives the presently highest spatial resolution available 
for polarization observations, i.e., 0\farcs0455 per pixel. Unfortunately, the pulsar and 
its nebula were placed mainly in the vignetted area of the PC chip. Since the part 
of the nebula was never more than 4\asec\ into the bad area, the loss of light 
was not so big. However, this made calibrations slightly more cumbersome. 

Due to the long exposures (600\,s) each frame exhibits thousands of cosmic 
ray (CR) hits. Most of these we removed by temporal median filtering of 
the three images, and the remaining hits are removed by {\it la-cosmic} which is
included in the IDL software. 
The images were taken at the roll angles 0, 45, 90 and 135 degrees, and to 
align the images three of them must be rotated and shifted. The de-rotation 
of 45 and 135 degrees requires interpolation, and so do the shifts. We used 
a cubic spline interpolation scheme and even though we succeed in overlapping 
the images to an accuracy of 10 milliarcseconds, it is clear that the 
resulting PSF (point spread function) for point sources may be affected. 
This means that we must be careful in our interpretation of polarization 
close to point sources. On the other hand, the extended emission, which is 
the focus of the present investigation, is not affected by the interpolations.

In this set of observations, the main source of instrumental polarization is 
the WFPC2 pick-off mirror. We have followed the calibration scheme as 
outlined in {\it WFPC2 Polarization Calibration Theory} \citep{Biretta97} to 
calculate the Mueller matrix for 605\,nm, and as a result we get following 
relations \smallskip




I  = 1.1 I$_{obs}$ 

Q = 1.1 (Q$_{obs} +$ 0.065 I$_{obs}$)

U = 1.6 (U$_{obs}+$ 0.034 I$_{obs}$), 

\smallskip

\noindent where I, Q and U are standard Stokes parameters, and I$_{obs}$,
Q$_{obs}$ and U$_{obs}$ are Stokes parameters before correction for
instrumental polarization. Test measurements of stars in the field show
a small tendency of a remaining 
systematic polarization of $\sim 2$\%, which is a bit more than 
expected. According to a detailed study by Jeffery (1991) of SN 1987A, 
which is located quite close to \psr, the foreground 
polarization is only $\sim 0.4$\%, so the $\sim 2$\% systemic uncertainty we
find probably also represents the uncertainty of our results for 0540. This is
small in comparison to the observed polarization which is typically 
10 $-$ 30\%, and therefore does not affect our results noticeably.

\subsection{Astrometric referencing}\label{astrom}

Astrometric referencing of the HST/WFPC2 images was performed based on  
the positions of the astrometric standards selected from the USNO-B1 
astrometric catalog\footnotemark\footnotetext{\tiny{USNO-B1 is currently
incorporated into the Naval Observatory Merged Astrometric Data set
(NOMAD) which combines astrometric and photometric information from 
Hipparcos, Tycho-2, UCAC, Yellow-Blue6, USNO-B, and the 2MASS, 
\href{www.nofs.navy.mil/data/fchpix}{www.nofs.navy.mil/data/fchpix}}}. 
A dozen of the USNO-B1 reference objects can be identified within the PC2 
chip FOV. Their 1$\sigma$ coordinate uncertainties vary between 50 and 900 
{\sl mas} with an {\sl rms} value of about 330 {\sl mas}. Pixel coordinates 
of the standards were derived using of the IRAF task {\it imcenter} with an 
accuracy of 0.05$-$0.1 PC2 pixel size (0\farcs042). The IRAF tasks 
{\sl ccmap/cctran} were applied for the astrometric transformation of the 
images. We consequently discarded standards which are either over saturated 
or have relatively large catalog and image positional uncertainties. 
As a result, only seven stars were selected for the final astrometric fit.
Formal {\sl rms} uncertainties  of the fit were best for the F814W filter 
$\Delta$RA$\la$0\farcs32 and $\Delta$Dec$\la$0\farcs43,  
and the fit residuals are $\la$0\farcs6, which is compatible with the 
maximum catalog position uncertainty of the finally selected standards 
(0\farcs33). The rest PC2 images were referenced to the F814W image  with 
the accuracy  better than 0.1 (0\farcs004) of the PC2 pixel size using 
several unsaturated stars from the PWN neighbourhood.     
Finally, the  first F547M images obtained in 1992 with the WFPC at a worse 
spatial resolution (pixel size 0\farcs1) were aligned to the WFPC2 images 
with the accuracy of about 0\farcs02.  

Inspection of the Chandra/HRC and ACIS images obtained in 2000 and 
2006  showed that within a nominal  Chandra  pointing accuracy of $\la$0.5  
arcsecond they are in a perfect agreement with the referenced optical images 
both by the pulsar coordinates and the orientation of the PWN. For the HRC 
observations of 1999 we found a large formal systematic shift of the pulsar 
coordinates form their correct values by 22.1 and -84.02 pixels of the CCD 
x and y coordinates, respectively.  After the correction by this shift   
the object position  and the PWN orientation are in agreement with those at
the other X-ray and optical images obtained at different epochs.  
 
All this allows us to compare the optical and X-ray structures of the PWN 
at a subarcsecond accuracy level. To get deeper X-ray images of 2000 and 2006   
the respective multiple data sets were combined making use of the 
{\sf merge-all v3.6 script} of the {\sf CIAO tool}.

\subsection{Wavelet filtering of the HST/WFPC2 images}\label{Wavel}

\subsubsection{Wavelet filtering}
In order to study the structure of the PWN in greater detail, a filtering 
has been applied to the image of the field around 
PSR B0540-69.3 (\cite{Starck98}). This method is 
based on a multi-scale transform of the image, the isotropic undecimated 
wavelet transform (also named ``\`a trous'' wavelet transform) that produces 
a set of bands at different scales, each band having the same number of pixels 
as the image. The original image $C_0$ can be expressed as the sum of all these 
wavelet scales ${w_j}$ and a smoothed version of the image $C_p$:

\begin{eqnarray}
C_0(x,y) &=& C_p(x,y) + \sum_{j=1}^p w_j(x,y)
\label{wave}
\end{eqnarray}
where $p$ is the number of scales used in the decomposition.

The filtering is obtained by applying a Hard-Thresholding to detect at each 
scale the significant coefficients :
\begin{eqnarray}
\tilde{w}_j(x,y) =  \begin{array}{ll}  w_j(x,y) & \textrm{ if  } w_j(x,y) \textrm{ is significant }\\
0 & \textrm{ if  } w_j(x,y) \textrm{ is not significant }  \end{array} 
\label{hard}
\end{eqnarray}

Assuming a stationary Gaussian noise, $w_j(x,y)$ has to be compared 
to $k \sigma_j$ :
\begin{eqnarray}
\label{signi}
\textrm{ if } |w_j(x,y)| &\ge& k \sigma_j \textrm{ then } w_j(x,y) \textrm{ is significant }\\ 
\textrm{ if } |w_j(x,y)| &<& k \sigma_j \textrm{ then } w_j(x,y) \textrm{ is not significant } \nonumber
\end{eqnarray}
If $w_j(x,y)$ is small, it is not significant and could be due to noise. 
If $w_j(x,y)$ is large, it is significant.
The threshold $k \sigma_j$ is obtained by estimating the noise standard 
deviation at each scale $\sigma_j$ and by choosing a $k$ value 
(see \cite{Starck06} for more details). We use the FDR technique to set $k$ in 
an adaptive way (\cite{Benjamini95}). The filtered image is obtained by the 
addition of the thresholded wavelet scales $\tilde{w}_j$ and the smooth 
version of the image $C_p$.

\subsubsection{Wavelet filtering to study the optical variability of the PWN}

To study the optical variability of the PWN, we have to focus on the diffuse 
emission surrounding the pulsar.
For this purpose, we have used a detection method to extract off the pulsar 
and other stars contained in the field. The detection method requires a 
multi-scale vision model defined in \cite{Bijaoui95,Starck06}, based on 
the ``\`a trous'' wavelet transform described previously. The multi-scale 
vision model (MVM) describes an object as a set of structures at different 
scales and a structure is a set of significant connected wavelet coefficients 
at the same scale $j$. Assuming a Gaussian noise, the definition of a 
significant wavelet coefficient is given by expression (\ref{signi}). A prior 
can be introduced to change slightly this definition. For instance, in order 
to extract off the pulsar and other stars, we can consider there is no 
interesting object larger than a given size. Then, we can force to 0, the 
wavelet coefficients larger than this size. Having detected all the interesting
objects in the field (e.g., pulsar and stars), they can be extracted by 
subtraction  (see \cite{Starck06} for more details). The remaining signal is 
the diffuse emission coming from the PWN. This is the emission we aim to study.


\section{Results}

\subsection{CTE effect and proper motion of \psr}\label{CTEpm}

\begin{figure}
\begin{center}
\includegraphics[height=80mm,angle=270, clip]{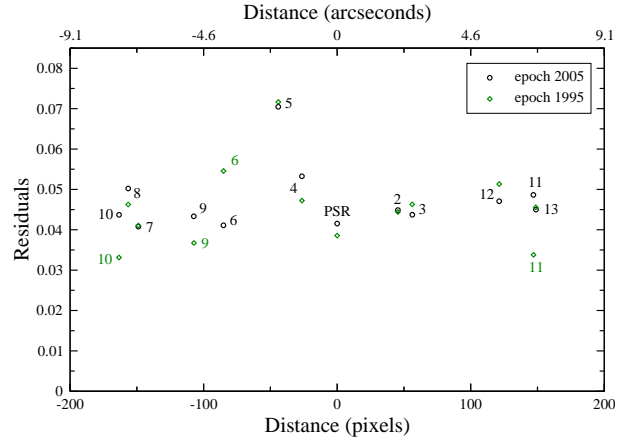}
\end{center}
\caption{Coordinate measurements of 13 stars in wavelet filtered F555W 
images for the 1995 and 2005 epochs (see text for further details.)
} \label{f:CTEtest}
\end{figure}

Modern techniques like wavelet filtering of images helps to bring out faint 
features and filters off traces of CTE. In order to investigate the importance 
of CTE effect on coordinate measurements or flux measurements we performed 
the following test on filtered and non-filtered images. We selected several
background stars close to \psr , including stars projected on the PWN. We used 
filtered and non-filtered F555W images for the 1995 and 2005 epochs. We measured 
coordinates on both sets of images using {\sf IRAF imcentroid}. Filtered 
and non-filtered images show no significant difference in accuracy within 
the image. A representative result obtained on filtered images is shown in 
Figure~\ref{f:CTEtest}, where we plot the distance between the pulsar and 13 stars 
with similar brightness as the filaments in the PWN, i.e., between magnitude 
$21-23$, versus the positional uncertainty of the stars. In general, fainter stars
produce higher residuals, but the numbers are low, and there is no clear trend 
that the 2005 image give higher residuals due to the CTE effect. We can therefore 
conclude that the CTE effect is unimportant for measurements of point sources, 
since the core of a star defines the position, which is unaffected by the CTE tail.

Main uncertainties arise when one aligns images from different epochs and 
sets of observations. The mean accuracy for this is about 0.1 of a
WFPC2/PC chip pixel (or 0\farcs005) (see~\ref{astrom}). At this accuracy 
level we do not find any significant displacement of the pulsar position 
for the considered  time base of 1999-2007. This is in agreement with the 
constraints on the pulsar proper motion recently reported by  
De Luca et al. (2007) using HST data between 1995-2005.

The situation becomes very different when we try to measure fluxes of point
sources, since part of the flux is in the CTE tail. We have tested different 
aperture sizes and find that the most optimal size is 10 WFPC2/PC chip pixels,
or $\sim$ 0\farcs5 . This aperture encapsulates a star of magnitude $21-23$  and 
its CTE tail. This aperture was chosen for most of our photometric measurements.

\begin{figure}
\begin{center}
\includegraphics[width=80mm, clip]{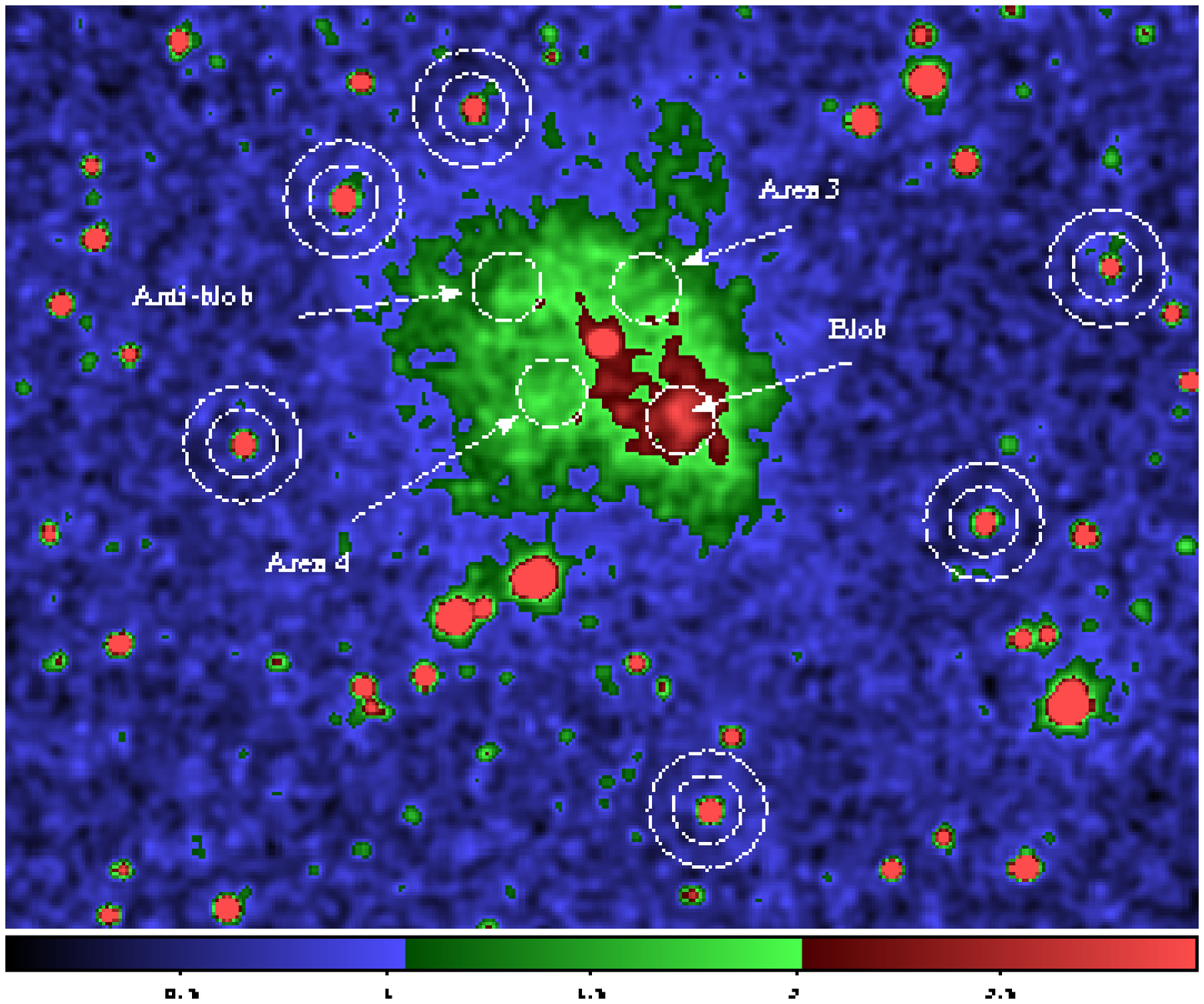}
\end{center}
\caption{\psr\ and its PWN as observed with HST/F547M in 1999. The
white circular apertures show the ``blob'' position in 1999, the ``anti-blob'',
``Area 3'' and ``Area 4'', as well as 6 references stars (see text for more
details). Some stars projected on the PWN were subtracted off. 
}\label{f:4areasbkg}
\end{figure}

\subsection{Optical photometry}\label{optphot}

To study spectral properties of the PWN we performed detailed photometry of 
selected areas in all optical continuum filters for 1999 and 2007.

We selected several areas in the PWN to perform photometry using the optical 
broad band filters. The aperture radius was 10 HST pixels, and the aperture was
centered on the maximum of flux of the ``blob''. Following \citep{DeLuca07}, we
therefore centered on different areas for the 1999 and 2005 images. 
However, for the 2007 epoch, there is no obvious dominant feature, and the 
aperture center corresponds to the assumed position of the ``blob'', had it continued 
to move with 0.04 times the speed of light as suggested by \citep{DeLuca07}.
An aperture of the same size was also placed on the apposite side of the pulsar
position along the major symmetry axis of the PWN and at the same distance away
from the pulsar as the ``blob''. We call this the ``anti-blob'' aperture 
(see Figure~\ref{f:4areasbkg}). 

\input{table-optf.tex}
We also selected two 10 HST pixel apertures along the minor symmetry 
axis of the PWN, symmetrically on both sides of the pulsar position, and 
name these ``Area 3''  and ``Area 4''. Thanks to the high spatial resolution 
of the HST/WFPC2/PC chip (0\farcs0455 per pixel), we were able to measure the
optical fluxes in these areas. 
For the optical background estimate 
for these areas we used annulus apertures of 6 HST pixels around 6 selected stars 
(see Figure~\ref{f:4areasbkg}). 


\begin{figure}
\begin{center}
\includegraphics[height=70mm,angle=270, clip]{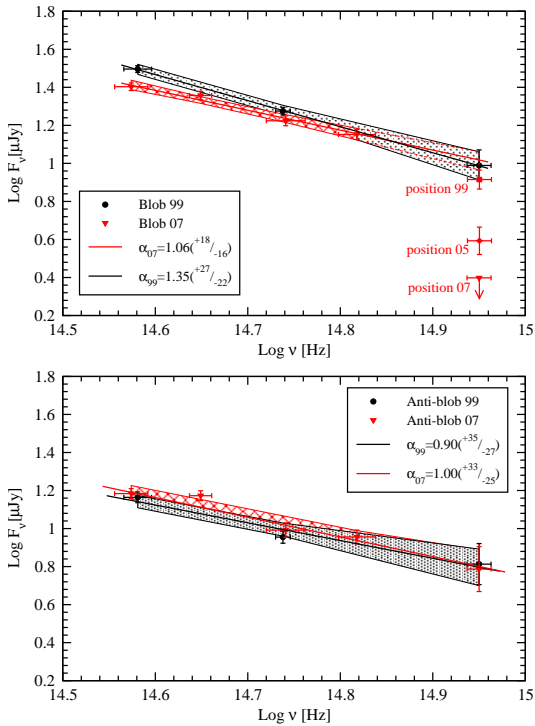}
\end{center}
\caption{Broad band optical spectra for the selected areas shown in
Figure~\ref{f:4areasbkg}. In the top panel we show the ``blob" spectrum in
red for 2007 and black for 1999. For the 2007 ``blob" spectrum we have applied
a $0\farcs4$ shift for the aperture, corresponding to a 0.04$c$ motion of the ``blob",
as suggested by \citep{DeLuca07}. For the spectral fit of this spectrum, we have 
omitted the F336W data (see text). The sensitivity of the F336W to where the
aperture is placed is shown in the plot. Data in other bands are less sensitive (see
Sect. 3.6). For the anti-``blob" we did not apply any shift in aperture position
between 1999 and 2007.
}\label{f:4areaSp}
\end{figure}


Photometry for all selected apertures was performed using {\sf IRAF/digiphot/phot}. 
The results are presented in Table~\ref{optical:fits} 
and Figure~\ref{f:4areaSp}. ``Area 3'' 
and ``Area 4'' show slight brightening from epoch 1999 to 2007, but with similar 
spectral index, $\alpha$, defined as $F_{\nu} \propto \nu^{-\alpha}$. The area
``anti-blob'' does not show any change within errors in flux and spectral index 
between the two epochs.

The most interesting behaviour is displayed by the ``blob''. First of all, the flux of the
``blob'' became significantly lower in 2007 compare to 1999. On other hand, the
spectral index became shallower in 2007 (see Table~\ref{optical:fits}), 
corresponding to a harder electron spectrum. There are two possible hypotheses:
either the ``blob'' moves with $\sim 0.04 c$ times the speed of light, as suggested by
De Luca et al. (2007) when they compared 1995, 1999 and 2005 images, or it fades away. 
We studied both these hypotheses. 

The ``blob'' was in its brightest phase in 1999. The aperture was centered on the
maximum of the flux within the area. In later epochs we placed the aperture on 
expected positions taking into account an assumed speed of $0.04 c$. For the F336W
image in 2007 the expected position of the ``blob'' is outside the visible PWN.
Since the PWN has quite steep power law, and since the extinction is higher in the blue,
we only have an upper limit on the flux of the ``blob'' in the F336W band (see
Table~\ref{optical:fits} and Figure~\ref{f:4areaSp}). This spectral point was not included
in calculations of the power law for the ``blob'' in 2007. All power law calculations 
were done using the algorithm described in Serafimovich et al. (2004). Fluxes in 
F555W and F675W are contaminated by emission lines of [O~III]$\wll4959,5007$ 
(\cite{Morse06}) and [S~II]$\wll6716,6731$, respectively, which could add some
systematic error to the continuum flux for those data entries in Table~\ref{optical:fits} 
and Figure~\ref{f:4areaSp}.

\subsection{Polarization of the PWN of \psr}\label{polarizD}


First optical polarization measurements for 0540 were performed by \citet{Chanan90}
using the CTIO 4m telescope. Chanan \& Helfand found a linear optical (V band) polarization, 
integrated over the PWN, which was $ P= 5.6\% \pm 1.0 \%$, oriented at a position angle
(eastward from north) $79\degs \pm 5\degs $. The time integrated observations with a spatial 
resolution of $0\farcs6$ per pixel did not allow them to detect any fine details. The first 
polarimetric phase-resolved observations of the \psr\ reported by \citep{Middleditch87}
gave only an upper limit to the polarization.

A few years later, time integrated optical polarization observations were performed 
by \citet{Wagner00} using the 8m VLT telescope. Wagner \& Seifert found up to 20\%  of linear 
polarization at the rim of the diffuse nebulosity, and 5\% for the pulsar.
In this case, the spatial resolution was $0\farcs2$, but the seeing conditions 
(from $0\farcs4$ to $1\asec$) could not bring out any detailed structures. 

\begin{figure}
\begin{center}
\includegraphics[width=87mm,clip=]{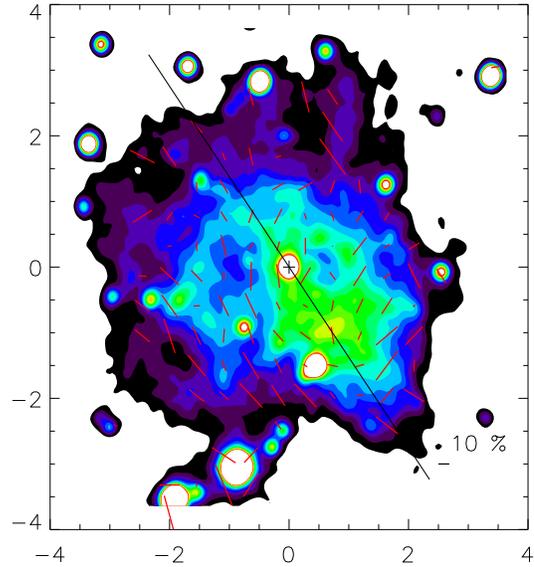} 
\end{center}
\caption{Continuum intensity (Stokes $I$) image at 602 nm of the PWN centered on \psr\ 
with overlaid linear polarization vectors. The size of the vector is degree of polarization in 
percents, while orientation of a vector is the position angle of linear polarization.  A 
horizontal tick
mark to the lower right shows 10\% degree of polarization. The major axis of the nebula, 
used for Figs.~\ref{polSlice:2007}, is marked by a solid line going from north-east to south-west, 
cutting across the pulsar and the ``blob" seen here in yellow.  
The x and y axes show distances north and west of the pulsar in arcseconds. 
}
\label{polF:2007}
\end{figure}

Polarization of the 0540 PWN has also been studied at 3.5, 6 and 20 cm in the radio, 
taking into account the Faraday rotation effect \citep{Dickel02}. Dickel et al. obtained 
a total fractional polarization of 20\% at 3.5 cm, 8\% at 6 cm and 4.5\% at 20 cm. 
They conclude that this large change in polarization with wavelength might be a result 
of high depolarization. The orientation is consistent with the optical value as 
reported by \citet{Chanan90}. \citet{Dickel02} noted that there is a gradient 
in rotation measure, which depends on magnetic field strength and electron number 
density, from northeast to southwest. 

The latest optical polarization observations of 0540 (see description in Section~\ref{polariz}) 
are compatible with earlier results, but bring out many new interesting details due to
the very high spatial resolution of HST/WFPC2. Figure~\ref{polF:2007} shows a continuum 
intensity image of the PWN with overlaid linear polarization vectors. Polarization 
vectors represent the electric field vector, $E$, which is perpendicular to the magnetic field. 
 The length of the vectors shows the degree of polarization.
The overall picture is quite complicated. As can be seen in Figure~\ref{polF:2007}, 
the outer parts of the PWN have high polarization, on the order of 30\% $-$
40\%. 
A similar behaviour has been seen for the Crab PWN 
(see \citet{Hester08} and references therein). 

The pulsar itself has low polarization, i.e. 5\% $\pm$ 2\%. This is consistent 
with earlier observations, but is in stark contrast to recently reported results of 
16\% $\pm$ 4\% by \citep{Mignani10}, using the same data set as we have used. The 
difference might originate from the difficulties of separating the pulsar from the nebula, which 
has high polarization. Another source of the different results might be different data 
reduction procedures (see section~\ref{polariz}), in particular the correction for the 
instrumental polarization.

\begin{figure}
\begin{center}
\includegraphics[width=78mm,clip=]{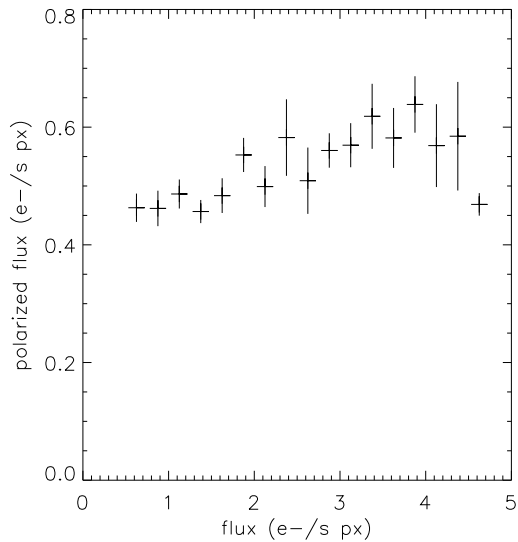} 
\end{center}
\caption{Polarized flux (Stokes $\sqrt{Q^2+U^2}$~) as a function of optical brightness 
(Stokes $I$) over the whole PWN in 2007. Note the weak correlation between polarized flux 
and intensity. The ``blob" area is the region with the highest intensity and smallest
error bar in polarized flux.
}
\label{polFlux:2007}
\end{figure}

The brightest parts of the 0540 PWN in optical continuum are also in general the 
brightest in polarized light. It is highlighted by the correlation in Figure~\ref{polFlux:2007}. 
This is perhaps not so surprising since the nebula emits synchrotron radiation, 
which can be highly polarized for ordered magnetic fields.
As can be seen in 
Figure~\ref{polF:2007}, the vector $E$ is perpendicular to, or at least not aligned with the major 
symmetry axis in the bright green area.
Such flow of charged particles, which could be a jet, may excite ejecta filaments and 
perhaps cause shock activity in the ``blob'' area (see section~\ref{morph}).
The possibility of a jet along the major axis was briefly mentioned by \citet{DeLuca07}. 
 However, what is most surprising in Figure~\ref{polFlux:2007} is how weak the correlation
is between optical continuum flux and polarized flux. The figure suggests that there is a general
smooth polarized flux in the nebula, on top of which there is more spatial variation in the non-polarized
flux. The ``blob" area stands out as the region 
with the highest intensity and smallest error bar in polarized flux. Its polarized flux
is slightly lower than other bright areas in the PWN, which may suggest less ordered 
magnetic field in the ``blob region".

\begin{figure}
\begin{center}
\includegraphics[width=80mm,clip=]{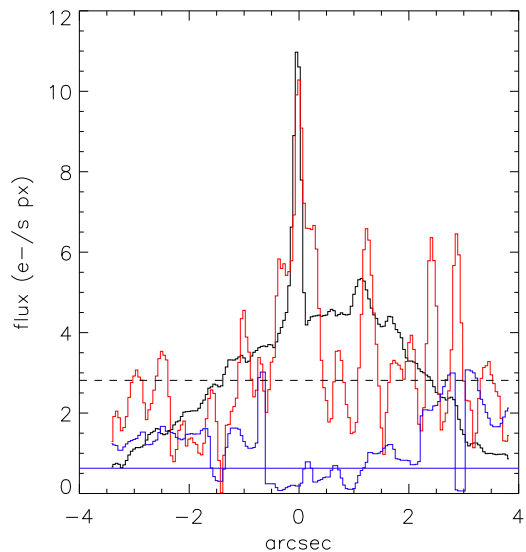} 
\end{center}
\caption{Flux (Stokes $I$), polarized flux (Stokes $\sqrt{Q^2+U^2}$~) and polarization angle 
along the major symmetry axis in 2007  (shown in Fig.~\ref{polF:2007}). The black curve is the flux, 
the red curve is the polarized flux multiplied by 10,  and the dashed line is the $3\sigma$ confidence level for 
the polarized flux. The blue horizontal straight line is the position angle of the major axis in Fig.~\ref{polF:2007}, 
and the blue curve indicates the polarization angle, $\theta$, in radians relative to noth (i.e, the position angle of 
the $E$ vector). Note how $\theta$ spans its full $0-\pi$ range.
The pulsar stands out as a peak in the flux and polarized flux at zero arcseconds, and the
peaks in flux and polarized flux at $\sim 1\farcs2$ correspond roughly to the ``blob" position in
the 1999 and 2000 optical continuum and X-ray images, respectively (see  
Figs.~\ref{ima-prof-547M.99} and~\ref{ima-prof-HRC-I.00} ).
}
\label{polSlice:2007}
\end{figure}

%
To explore the polarization properties more in detail along the major symmetry axis,
we have studied a $1\asec$ wide slice, centered on and along the major axis. We will 
later also use a similar slice for X-ray projections and the non-polarized optical data 
(see below). Figure~\ref{polSlice:2007} shows flux (black), polarized flux (red) and 
polarization angle (blue). We also show the position angle of the major axis (blue horizontal
line). The largest maximum 
(around $1\farcs2$) to the right of pulsar is close to the ``blob" position in 1999. 
We can see that the polarized flux jumps by a factor of $\sim 3$ at this position, with a 
small shift by $\sim 0\farcs1$ further away from the pulsar for the polarized flux
compared to the continuum intensity. 
At the same time, the polarization angle  swings $\sim 60^\circ$ going from the inside of the
blob to where the polarized flux peaks. The alignment of the $E$ field and the major axis is
$\sim 20^\circ$ at the position of peak intensity. This is close to what is observed for presumed
shocks in extragalactic jets, where the $E$ vector abruptly swings from perpendicular to the jet, 
to parallel, at jet knots. We see a similar behavior 
on the opposite side of the pulsar at about $1\farcs0$, although the structure is less clear. 
That the position angle of the polarization vector changes just before the polarized flux 
reaches a maximum at the ``blob'' position, and perhaps close to the  ``antiblob" position, 
could point to shock activity (present or past) in those regions. 

We can also see two maxima in polarized
flux well above $3\sigma$ on the same side as the ``blob'', but at $\sim 2\farcs3$ and 
$\sim 2\farcs8$ away from the pulsar. For the one at $\sim 2\farcs3$, the position
angle swings away from being parallel to the symmetry axis to being perpendicular, 
and at $\sim 2\farcs8$ the electric vector is parallel to that between the pulsar and the ``blob''.
Note that the polarized flux is almost as bright in those regions as it is at the ``blob'' position. 
We see a weak, but similar trend, also on the opposite side of the pulsar, outside the 
``antiblob'' position.

\subsection{X-ray counterpart of the ``blob''}\label{projections}

\begin{figure*}
\setlength{\unitlength}{1mm}
\resizebox{12cm}{!}{
\begin{picture}(120,120)(20,45)
\put (0,165) {\includegraphics[width=70mm, angle=-90,clip=]{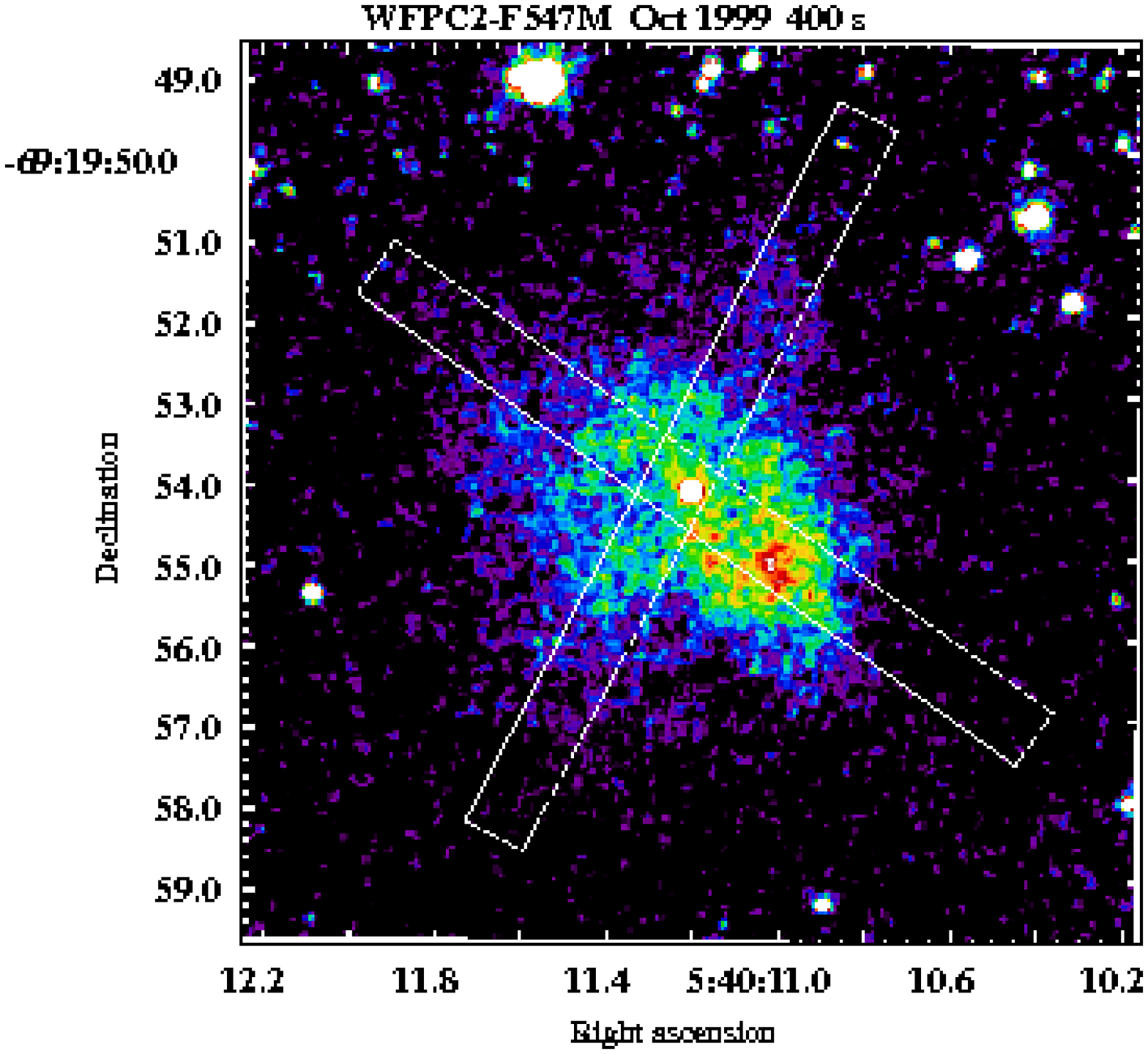}} 
\put (85,163) {\includegraphics[width=40mm, angle=180,clip=]{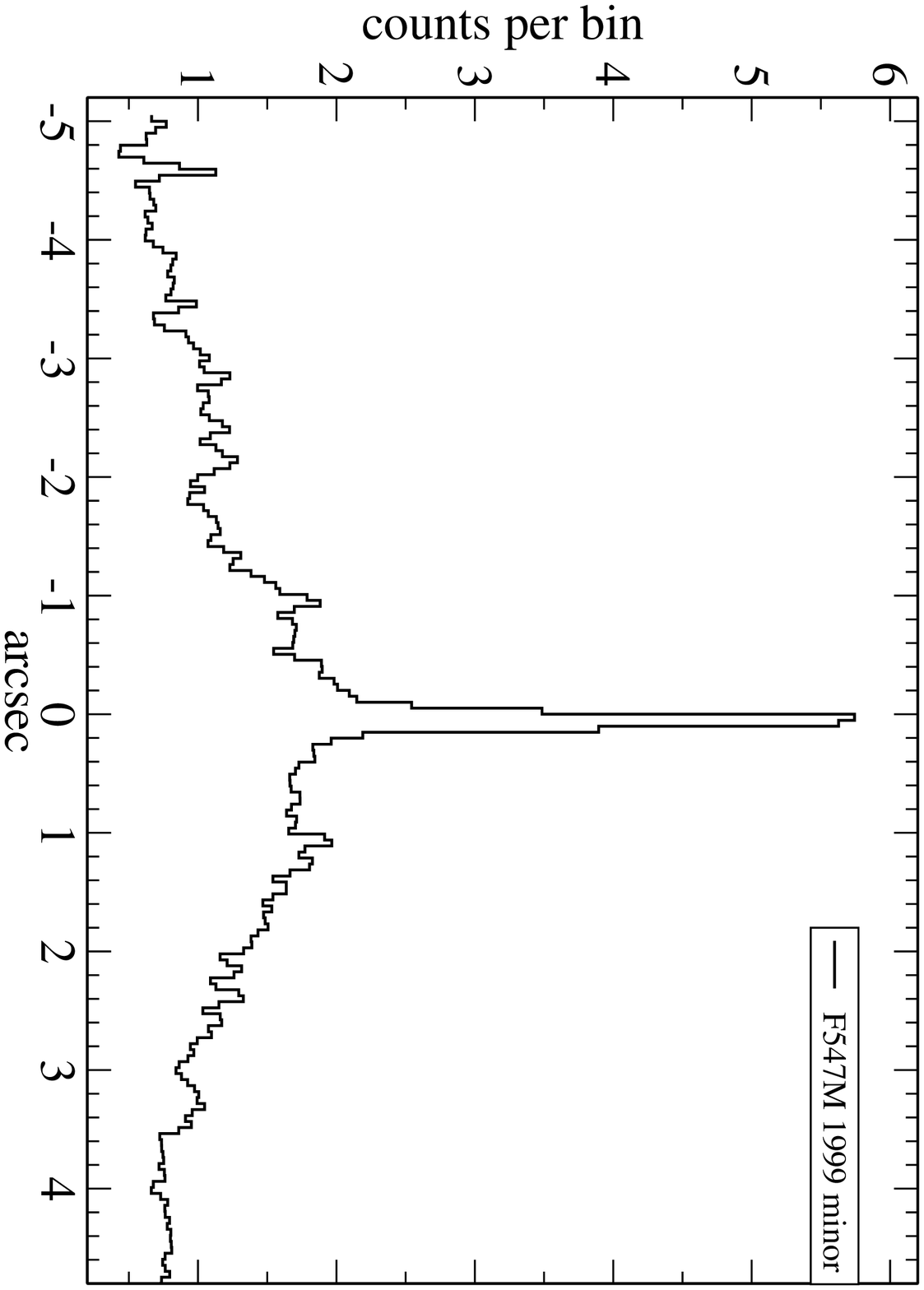}} 
\put (16,93) {\includegraphics[width=40mm, angle=-90,clip=]{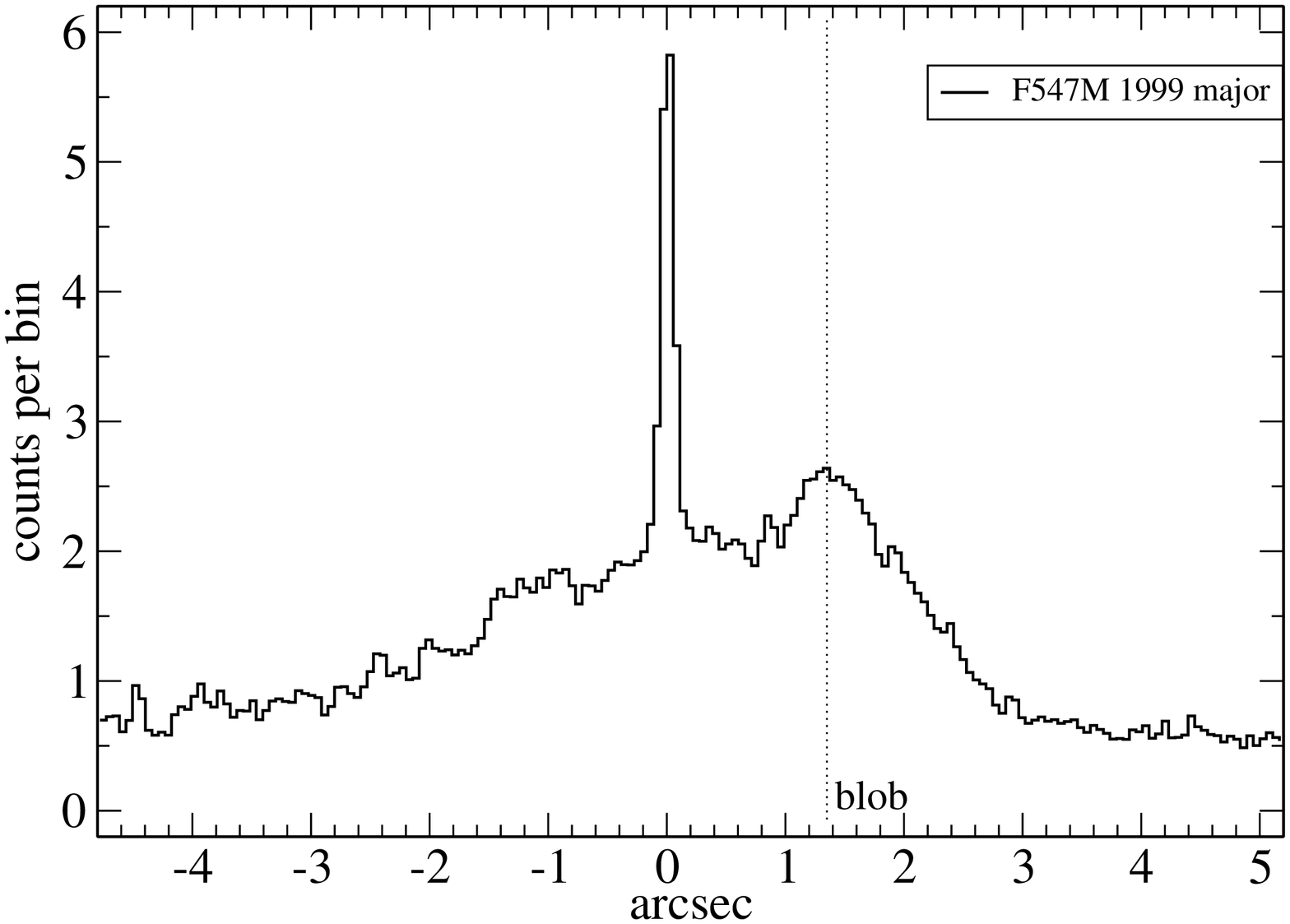}} 
\put (80,100) {\parbox[t]{70mm}{\caption{{\sl Top-left:} 11\asec$\times$11\asec~optical 
image of the 0540 pulsar+PWN system taken in October 1999 with the HST/WFPC2 in 
the F547M band. Positions of two slices with the sizes of 0\farcs8$\times$10\asec 
are marked by white lines. They were used for the extraction of the 1D spatial 
profiles  along the minor (NW-SE) and major (NE-SW) axes of the system shown in 
the {\sl top-right } and {\sl bottom} panels, respectively. Coordinate origins 
of the spatial axis in the profile plots coincide with the position of the pulsar 
visible as a bright point source near the PWN center. The position of the  bright 
PWN ``blob'' seen southwest of the pulsar is marked by a vertical dotted line at 
the {\sl bottom} panel. 
}\label{ima-prof-547M.99}}}
\end{picture}}
\end{figure*}
\begin{figure*}
\setlength{\unitlength}{1mm}
\resizebox{12cm}{!}{
\begin{picture}(120,120)(20,45)
\put (0,165) {\includegraphics[width=70mm, angle=-90,clip=]{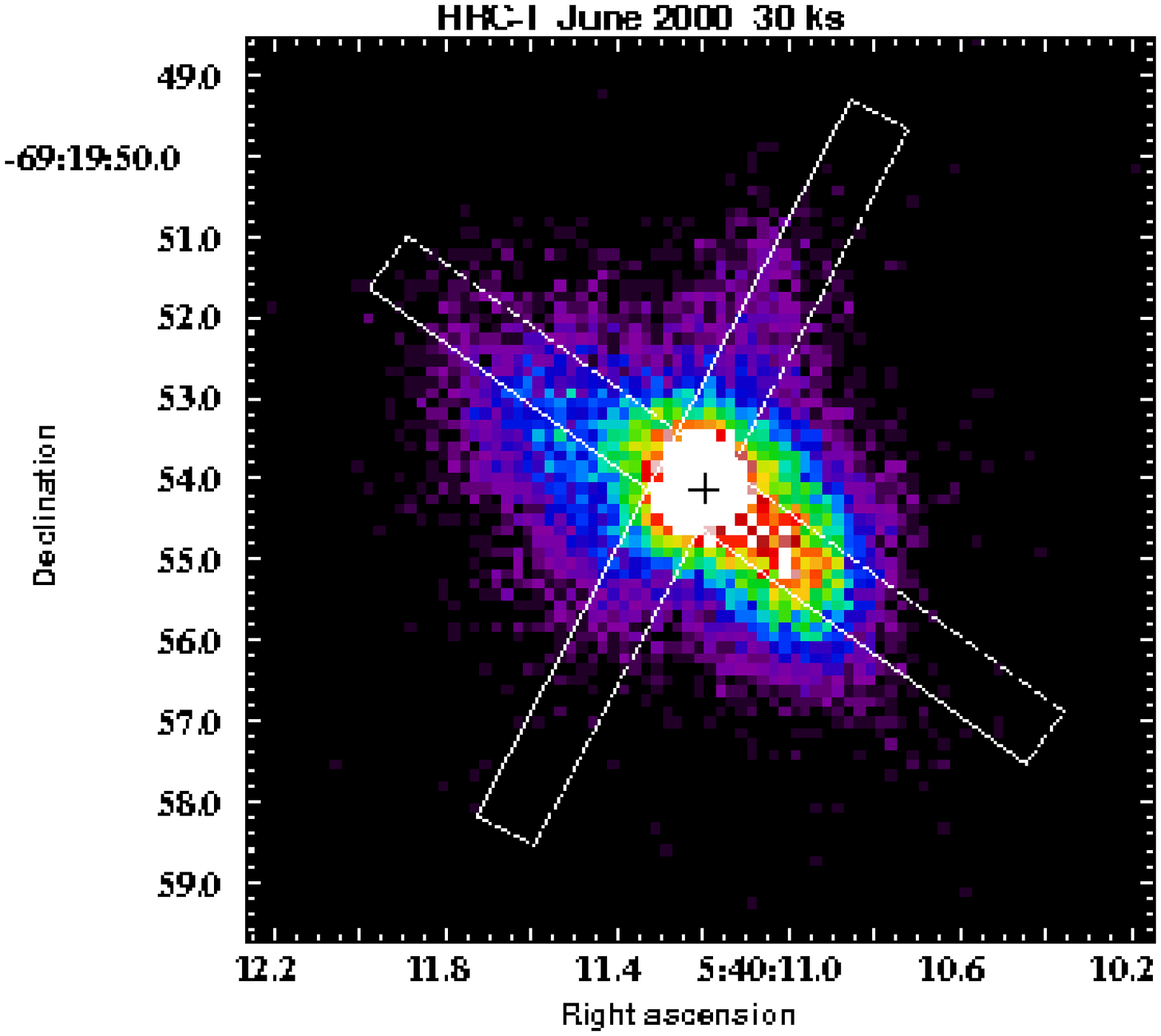}} 
\put (85,163) {\includegraphics[width=40mm, angle=180,clip=]{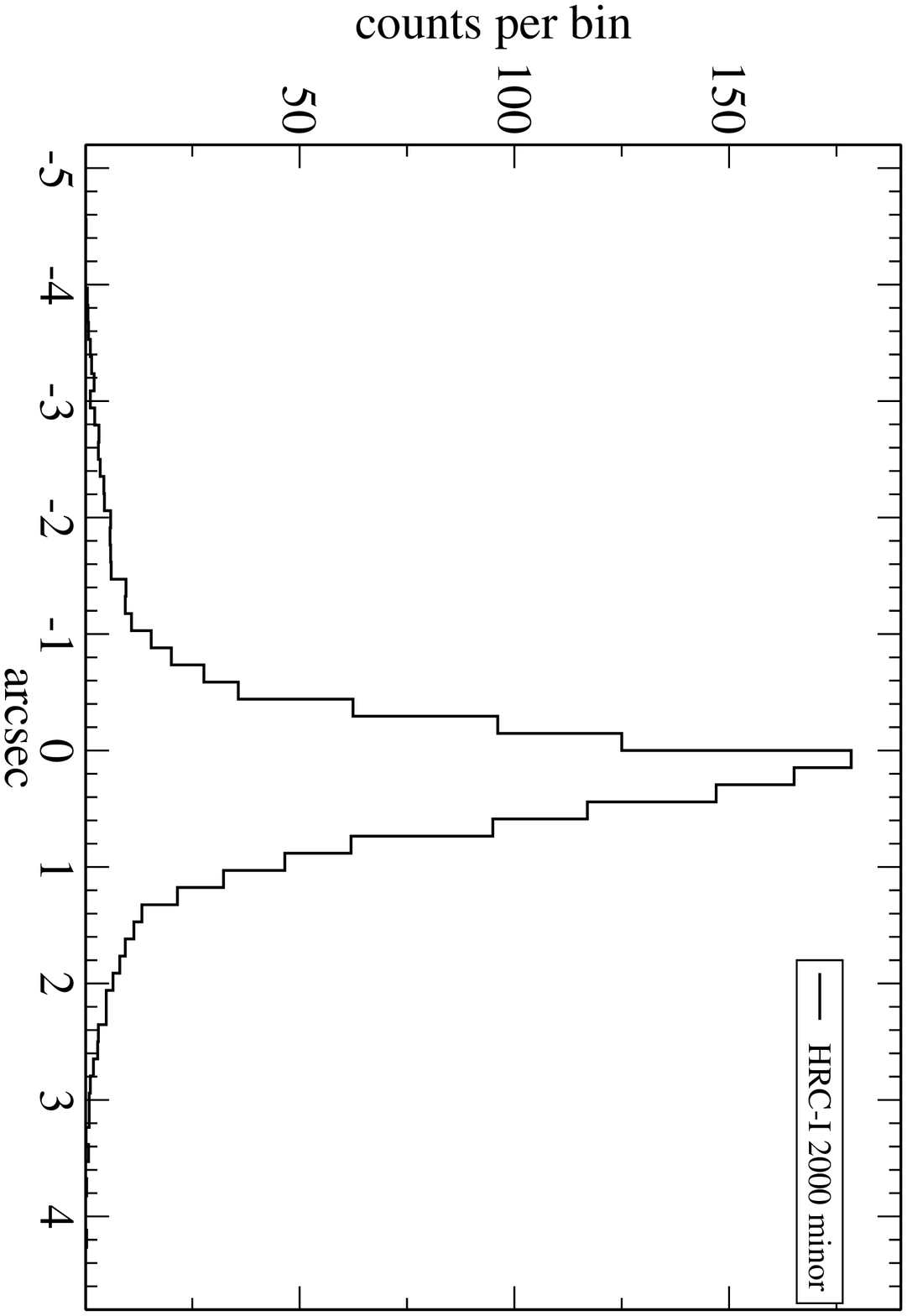}} 
\put (15,93) {\includegraphics[width=40mm, angle=-90,clip=]{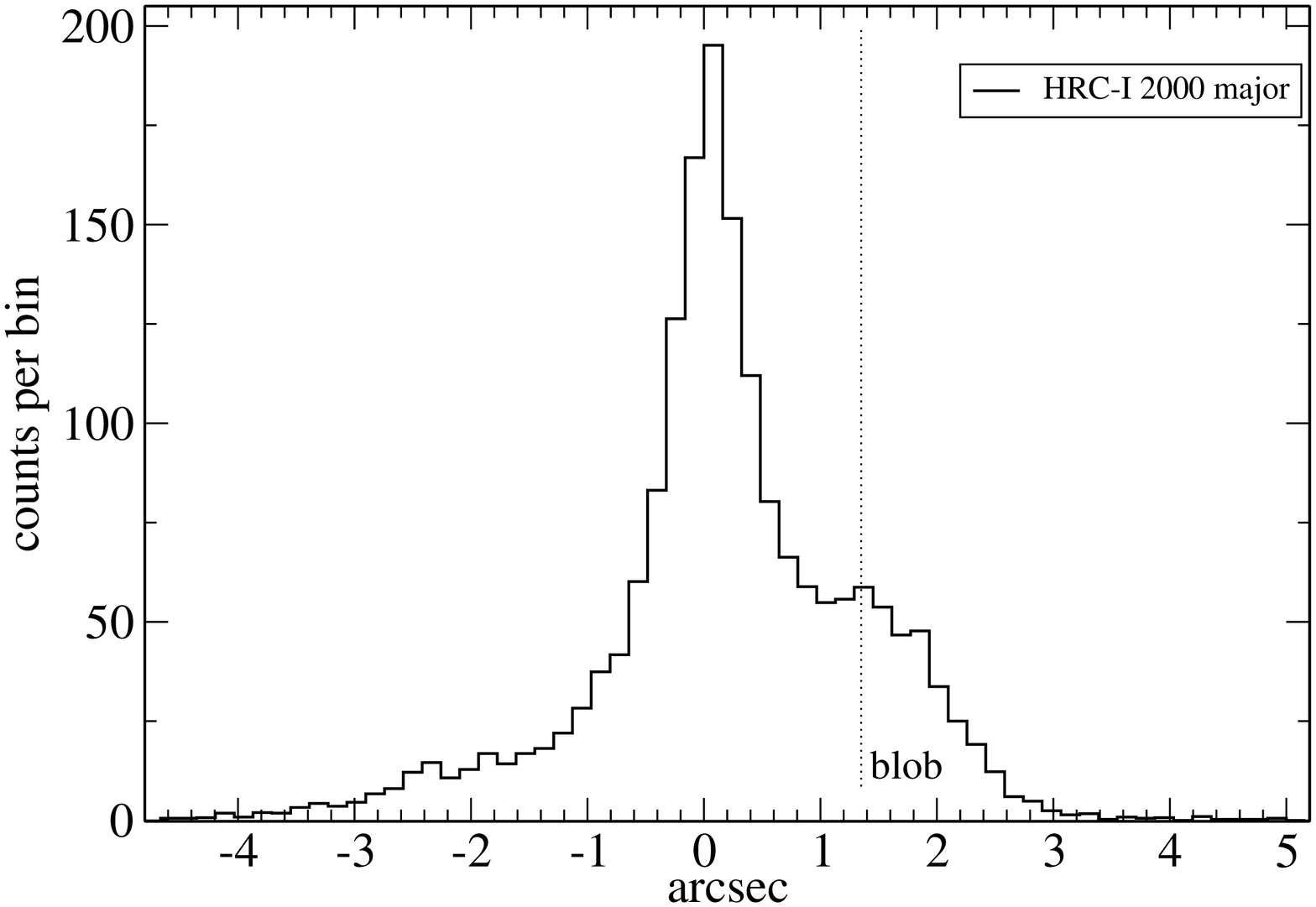}} 
\put (80,100) {\parbox[t]{70mm}{\caption{Same as in Figure~\ref{ima-prof-547M.99}.
but for the X-ray data taken in June 2000 with Chandra/HRC-I in the 0.2$-$10 
keV band. The slice positions and sizes are the same as in the optical image. 
The cross in the {\sl top-left } panel marks the optical position of the pulsar, 
while the vertical line at the {\sl bottom} panel shows the position of the PWN 
``blob'' in the optical in October 1999 (see also the left panel of Fig.~\ref{X-ray_index}
for the 2006 Chandra/ACIS data). As seen, despite the half year difference 
between the optical and X-ray observations, a likely X-ray counterpart of the 
optical ``blob'' can be seen at the same place as in the optical range. 
}\label{ima-prof-HRC-I.00}}}
\end{picture}}
\end{figure*}
 
To investigate how the X-ray emission correlates with the optical and polarization
data, we carefully aligned the optical and X-ray images as described in section~\ref{astrom}, 
and extracted intensity profiles along major and minor PWN symmetry axes. We chose 0\farcs8 
by 10\asec\ slit size to cover the ``blob'' in the best way, taking into account the resolution of 
different instruments. 

Figures~\ref{ima-prof-547M.99} and~\ref{ima-prof-HRC-I.00} show an optical image in
F547M taken in October 1999 and an X-ray image in HRC-I taken in June 2000, with 
their respective profiles. The position of the ``blob'' is marked by vertical dotted line 
at the bottom panel in Fig.~\ref{ima-prof-547M.99} (see also the left panel of Fig.~\ref{X-ray_index}
for the 2006 Chandra/ACIS data). The spatial profile clearly shows 
a maximum of flux from that area. As can be seen in the bottom panel in Fig.~\ref{ima-prof-HRC-I.00} 
the same optical coordinates correspond to a maximum of flux from the ``blob'' area also in 
X-rays. This clearly shows that the enhanced X-ray and optical radiation come from the same 
area.

We have performed the same investigation for the second epoch for which we have near-coincident
optical and X-ray data, i.e., for 2005/2006. The optical emission shows a change in morphology of the 
``blob'' as discussed later in the paper, while the X-ray emission does not show any change in morphology
to within the spatial resolution. Note that the worse spatial resolution of Chandra/ACIS compared to 
Chandra/HRC.

\subsection{ACIS-S  X-ray spectroscopy}\label{Xrayspec}

Kaaret et al. (2001) were the first to attempt a study of the X-ray spectral 
structure of  the 0540 PWN. They used the observation carried out on August 
26 1999 with the ACIS-I  in a  continuous-clocking (CC) mode. A one-dimensional
image oriented approximately along the major axis of the nebula including the pulsar 
was analyzed. Spectra extracted from  the western and eastern sides of 
the pulsar, excluding a 0\farcs7 region around the pulsar, showed similar 
photon spectral indices of $\alpha$=0.96$\pm$0.11 and  
1.12$\pm$0.14, respectively, for a fixed  $N_H$ column density of 
4.6$\times$10$^{21}$ cm$^{-2}$, assuming Milky Way (MW) composition of the absorbing gas. 
The observed flux was twice as bright from the western region than from the eastern. 
Later, utilizing the ACIS-S observation obtained on 
November 22-23 1999, Petre et al. (2007) measured the dependence of the PWN 
photon index on the radial distance from the pulsar. They extracted spectra  
from concentric elliptical annuli centered on the pulsar and aligned with 
the PWN major axis. A systematical softening of the nonthermal PWN emission 
with radius was found, as  is also observed in the case of the Crab PWN. However, 
the azimuthal averaging made in the Petre et al. study does not allow to reveal 
any possible asymmetry in the photon index spatial distribution, as suggested by  
the PWN morphology.
A strong pileup effect  
in the ACIS-S 1999 data set  precludes a relaible spectral analysis, owing to 
the conversion of two photons into a single event that has 
apparent energy equal to the sum of the photon energies, leading to 
significant spectral distortions and  a loss of photons.  


The new ACIS-S data listed in Table 2 allow us to perform a more detailed and reliable study of 
the spectral structure of the PWN owing to a significantly smaller pileup and much longer exposures. 
However, the spectra from the bright PWN regions, including the pulsar and the blob, are still distorted 
by a moderate pileup. This leads to artificial excesses of high energy events reducing the derived 
photon indices. For instance, a single absorbed power law fit is unacceptable for the pulsar and 
leads to an implausible photon spectral index of only half that in previous works.

To correct for that, the ACIS {\sl pileup model} was included. To check its applicability we first examined 
the spectrum of the pulsar, since it is a point-like object and the {\sl  pileup model} was developed 
for point-like sources. We extracted data from a 3$\times$3 pixel bin centered on the puslar. 
Trailing events are seen in the ACIS images as a faint east-north-south-west streak, emanating from 
the pulsar along the CCD readout direction, and is caused by photons detected from the pulsar 
during frame readouts. These were placed back to the pulsar position using the CIAO {\sl acisreadcorr} tool. 
The total number of the trailing events was estimated to be less than 1.5\%~of the total count number 
 from the pulsar region (with a count-rate of $\sim$0.5 counts s$^{-1}$). Backgrounds were taken 
from a 6\asec\ radius aperture with the coordinates RA=05:40:36.876 and Dec=-69:22:17.38 far 
away of the remnant.

The spectrum was fitted in the 0.3--10 keV range with an absorbed power-law convolved 
with the pileup model 
using standard XSPEC v12.3 tools. In this test for the absorbing column density, we assumed MW 
abundances as had been done in most previous studies of the object. We approximated the
MW abundances by the solar abundances of Anders \& Grevesse (1989). In general, the solar metallicity 
according to Anders \& Grevesse is $0.1-0.3$ dex higher than the Milky Way abundances adopted by
Wilms et al. (2000). However, the Anders \& Grevesse abundances are widely used, which makes
it easy to compare with previous results. As discussed below, our conclusions do not depend on
using Anders \& Grevesse (1989) for the Milky Way abundances.

The resulting spectral index for the pulsar, $\alpha$=0.9$\pm$0.06, is in excellent agreement 
with $\alpha$=0.92$\pm$0.11 for the rotation-phase-averaged ``pulsar all" spectrum of Kaaret et al. (2001) 
obtained using ACIS CC  (continuous-clocking) 
observations, and with the Petre et al. (2007) value of 0.92$\pm$0.25 obtained by extracting only pulsar
trailing events from the 1999 ACIS-S observation. In both these latter cases, the pileup is negligible. The fit also
provides a reasonable PSF fraction {\sl psfrac}~$\sim 1$ treated for the pileup, and a plausible {\sl probablility} 
$\sim$0.52 that adding a photon to an event yields a valid X-ray event. The column density, 
$N^{MW}_{\rm H}$=(5.0$\pm$0.01)$\times$10$^{21}$ cm$^{-2}$, was compatible with that of Kaaret et al. (2001) 
and Petre et al. (2007). This ensures us that the moderate pileup in the 2006 data-set can be reliably corrected for. 
The derived absorbed flux from the pulsar region in the 0.6--10 keV range
is $\sim$1.45$\times$10$^{-11}$ erg sm$^{-2}$ s$^{-1}$,
which is $\sim 20$\% lower than that of Kaaret et al. (2001). This may be explained by some 
flux contamination from the relatively bright neighboring nebular regions across the readout direction which 
are collapsed together with the pulsar emission in the 1D image in the continuous-clocking mode. It could also
be caused by a difference in assumed abundances for the Milky Way, i.e., if Kaaret et al. (2001) used
abundances with somewhat lower metal content than in Anders \& Grevesse (1989).
\begin{figure}
\begin{center}
\includegraphics[width=80mm, clip]{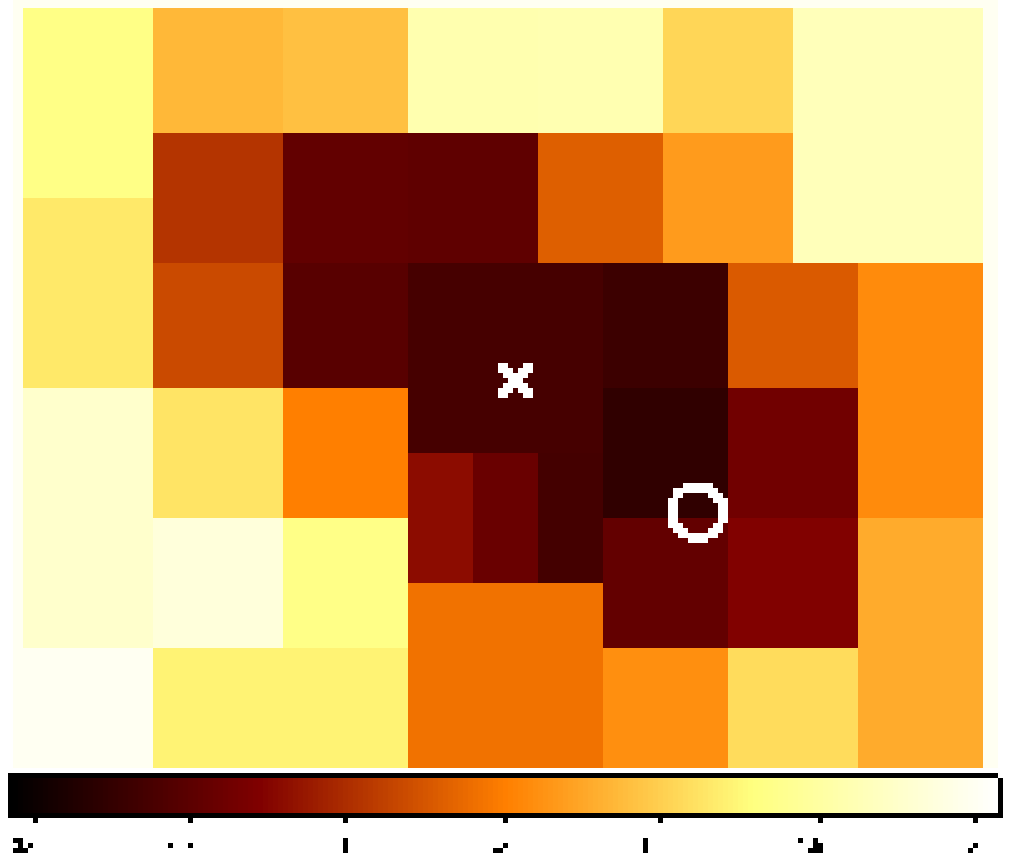}
\end{center}
\caption{X-ray spectral index variation map of PWN 0540. The color
coding represents values of the spectral index $\alpha_{\nu}$, 
which varies from 0.65 to 1.8. North is up and east to the left. The white cross is 
the pulsar optical position, and the white circle is the optical aperture centered 
on the ``blob'' position in 1999. Note that this region overlaps with the area which 
has the smallest spectral index,
$\alpha_{\nu} = 0.65\pm0.03$ (see Table ~\ref{xray:fits}), in X-rays. The distance
between the pulsar and the blob is $1\farcs4$.
}\label{f:indexGammaIma1}
\end{figure}

To continue our check, we also included a pileup model for the emission, integrated from large extended
 regions of the PWN, e.g., from the semi-elliptic apertures encapsulating the total emission from the left (eastern) 
 and right parts of the nebula along its major axis. For the fainter left part, where pileup is negligible, the model 
 only marginally improves the fit, while for the brighter right part the improvements are more significant. 
  After the pileup correction, the latter shows a harder spectrum than the former one, with the differences 
 in  $\alpha$ of $\sim 0.2$ and 
 in the observed fluxes by a factor of $\sim 2$. That is again fully consistent with what was published 
 by Kaaret et al. (2001, Table 2) based on the CC observations. This demonstrates that the new ACIS-S data 
 in combination with a pileup model can be used also for the spectral analysis of the extended PWN, 
 at least at a qualitative significance level.

As was discussed in Serafimovich et al. (2004), MW abundances are not the best to use for 0540.
To investigate that, we analysed the pulsar spectrum using different absorbing element abundances 
in the MW and LMC, splitting the total column density as  $N_{\rm H}$=$N^{MW}_{\rm H}$+$N^{LMC}_{\rm H}$. 
The difference in absorption between MW and LMC gas was shown in Serafimovich et al. (2004) 
for the pulsar/PWN system, and was then used by Park et al. (2009) in their study of outer regions of 
the SNR with the same data set. We fixed $N^{MW}_{\rm H}$ at the value of 
0.6$\times$10$^{21}$  cm$^{-2}$, while  $N^{LMC}_{\rm H}$ was found from the fit. 
We obtained $N^{LMC}_{\rm H}$ =(0.8$\pm$0.02)$\times$10$^{22}$ cm$^{-2}$, assuming the LMC 
abundances of Russel \& Dopita (1992). The resulting spectral index was $\alpha=0.7\pm$0.02, which is 
significantly harder than in previous works. The derived LMC column density is about 30\% - 50\%  higher 
than the upper limit discussed by Serafimovich et al. (2004), and the value 0.6$\times$10$^{22}$ cm$^{-2}$ 
obtained by Park et al. (2009). Fixing $N^{LMC}_{\rm H}$ at the latter level produces an unacceptable fit for 
a single power-law spectrum to be typical for all young pulsars detected in X-rays. 
 
As discussed at length in Serafimovich et al. (2004), the LMC column density is unlikely to
be much higher than 0.6$\times$10$^{22}$ cm$^{-2}$. To resolve this problem, we included the idea
of Serafimovich et al. (2004) that there could be a third absorbing component, namely that from the
ejecta of 0540. So, we fixed  $N^{MW}_{\rm H}$ and $N^{LMC}_{\rm H}$  at their 
reasonable values of 0.6$\times$10$^{21}$ and 0.6$\times$10$^{22}$ cm$^{-2}$, respectively, and included 
the third component, $N^{SNR}_{\rm H}$, assuming that it is produced by the SN ejecta. It has to be 
dominated by heavy elements, like He, C, O, Si, and Fe, which are found in optical spectra 
of the remnant (e.g., Serafimovich et al. 2005). Since 0540 is an oxygen-rich SNR we started with 
only including hydrogen+oxygen, whereby we obtained a better fit ($\chi^2$ near 1.1 per 
 d.o.f.) with $N^{SNR}_{\rm H}$ a few times 10$^{19}$ cm$^{-2}$ and O/H 
$\sim100$ times the solar  
 value of Anders \& Grevesse (1989), which is $8.5\times10^{-4}$, by number. The spectral index remained at 0.7.
 
The estimated amount of ``extra" oxygen is close to that derived for SN 1987A, which can be considered as a 
good model for an oxygen-rich event in the LMC. The model of Blinnikov et al. (2000) for SN 1987A was used
and expanded to the size of 0540 in Serafimovich et al. (2004). 
 The Blinnikov et al. model rests on the progenitor model calculated by Nomoto \& Hashimoto (1988) and 
Saio et al. (1988), which assumes that 14.7 $\Msun$ are ejected at the supernova 
explosion. The total mass on the main sequence is $23 \Msun$, but several solar masses are lost in stellar winds. 
The progenitor model, which is one-dimensional, is normally referred to as the 14E1 model (Shigeyama \& Nomoto 
1990). This model consists of well-defined shells of different elements, with the heavier elements toward the 
center. Blinnikov et al. (2000) used this model to calculate the broad-band filter emission from SN 1987A, but found 
that substantial mixing of the ejecta was needed to obtain good fits to the filter lightcurves. Figures 2 and 3 
in Blinnikov et al. (2000) show the preferred structure after mixing, and the unmixed model, respectively.
Including the same elements (H, He, C, O, Si and Fe) as in Serafimovich et al. (2004), the abundances relative 
to Anders \& Grevesse (1989) for  $N^{SNR}_{\rm H}$ were assumed to be He=5.75, C=32.5, O=151.58, 
Si=65.95, and Fe=50.47. Other elements were neglected since they probably do not contribute significantly 
to the absorption. As a result, we obtained $N^{SNR}_{\rm H}$ =(2.42$\pm$0.09)$\times$10$^{19}$ cm$^{-2}$,   
$\alpha$=0.745$\pm$0.014, and a normalization constant of (2.30$\pm$0.25)$\times$10$^{-3}$ photons cm$^{-2}$s$^{-1}$ 
keV$^{-1}$ at the reduced $\chi^2$=1.02 for 851 {\sl d.o.f}. The harder spectrum of the pulsar, as compared 
to previous works, is likely explained by effectively lower absorption than in case of pure MW element  
abundances of the absorbing matter. We note, that based on the current data quality and the fits statistics, we 
cannot justify with great confidence that the obtained splitting of the absorption in three parts is real, since for 
the pure MW and/or MW+LMC column densities the fits are also acceptable. However, the latter leads 
to an  $N^{LMC}_{\rm H}$ value which we believe is too high (see also the discussion in Serafimovich et al. 
2004). The main problem with the X-ray data is the pileup effect. Different pileup corrections for different element 
abundances cannot be excluded, although formally we got similar values of  {\sl psfrac}$\approx$1, 
and  {\sl probablility}$\approx$0.5, i.e., adding a photon to an event yields a valid X-ray event, for all three 
cases of the abundances models considered. Nevertheless, including absorption from the supernova ejecta
is physically reasonable. 
  
\input{table-Xray.tex}

Thus, based on the results from the spectral analysis of the pulsar emission, we fix the MW, LMC and SNR 
column densities at the above values in the following spectral analysis of the PWN. 
To study in detail the spatial spectral variation within the PWN, we divided its image 
into rectangular spatial bins. We used 2$\times$2 pixel bins for the bright central parts and 2$\times$4 
and/or 3$\times$4 bins for the faint outer regions to provide enough count statistics there. Background fluxes 
were taken from the same region as for the pulsar. The spectral data for all ACIS-S data-sets were grouped to 
provide a minimum of 15 counts per spectral bin for faint regions and of 25 for brighter ones. 
The total number of source counts varied from about 1000 for faint regions to about several tenths of 
thousands for the blob (with a total source count rate of $\sim$0.125 counts s$^{-1}$ 
in the 0.5--10 keV range). The background fluxes were about $2.3\times10^{-5}$ counts arcsec$^{-2}$ s$^{-1}$, which is 
insignificant compared to the source count rates. The spectra were fitted in 0.3--10 keV range with 
the absorbed power-law. The pileup corrections were significant only in some relatively bright regions 
in the right part of the PWN, including the ``blob". For the compact ``blob", the pileup model parameters were compatible 
with those of a point source (see above), while for other bright region, where the pileup is still  
significant, they are  less constrained.  The results of the fit for the blob/ant-blob regions are presented in the 
Table ~\ref{xray:fits}, and in Figure~\ref{f:indexGammaIma1} we show the spectral index map of the PWN, compiled from the
spatially resolved X-ray spectral analysis described above.

\begin{figure*}
\setlength{\unitlength}{1mm}
\resizebox{12.cm}{!}{
\begin{picture}(100,50)(0,0)
\put (50,50) {\includegraphics[width=45mm, angle=-90,clip=]{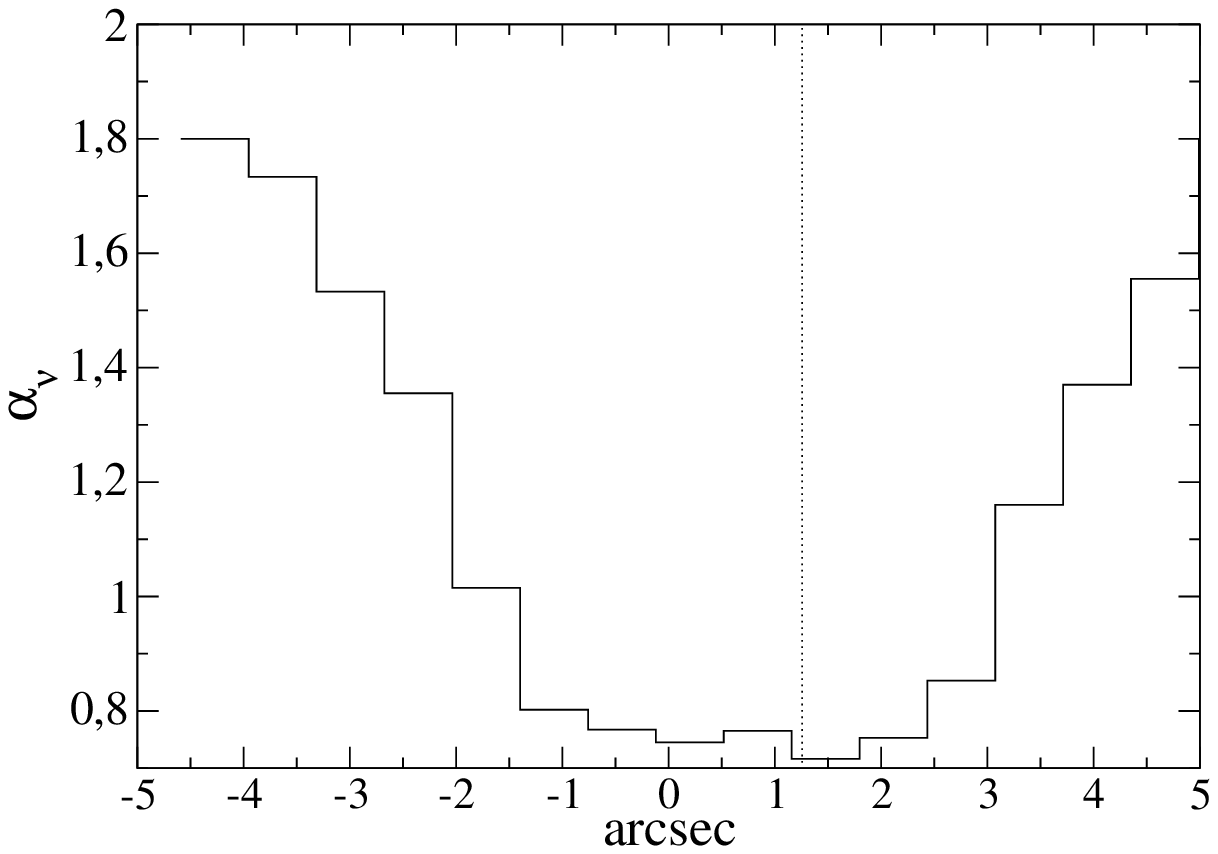}} 
\put (-20,50) {\includegraphics[width=45mm, angle=-90,clip=]{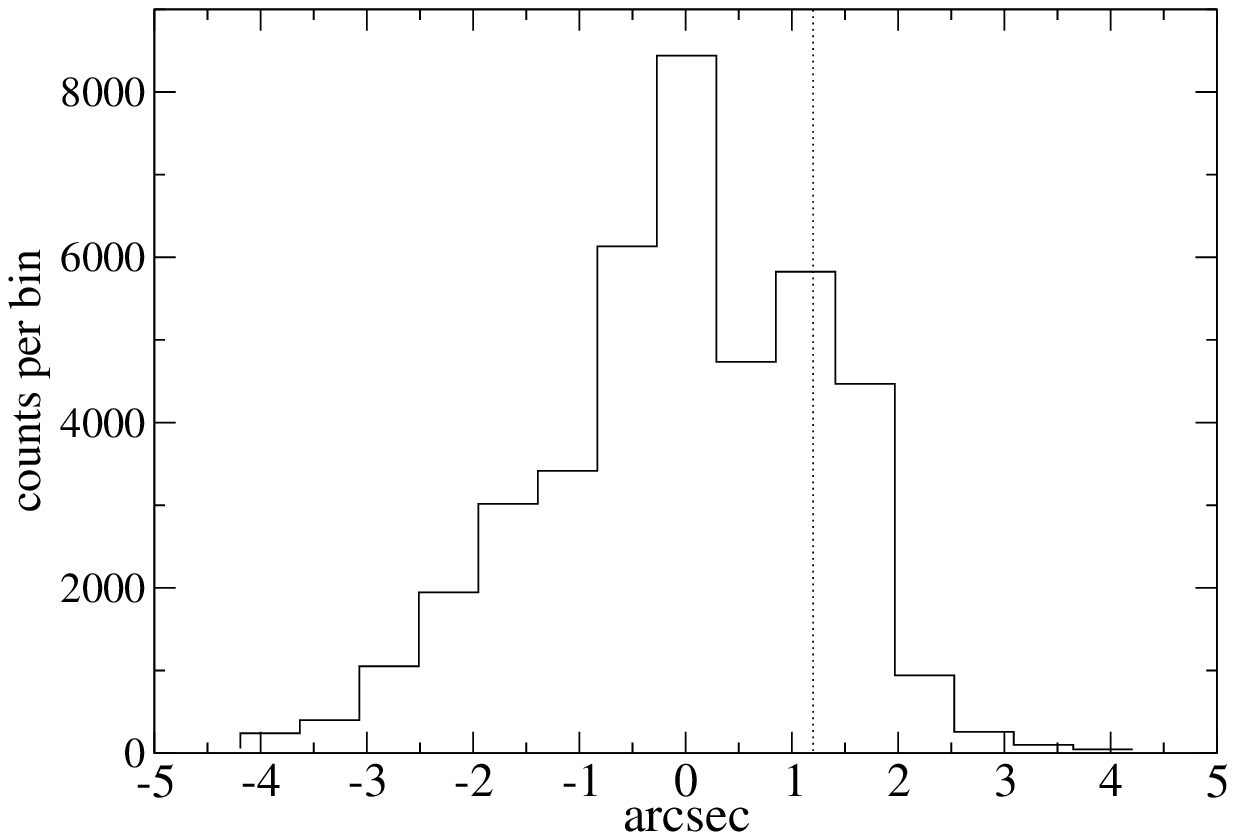}} 
\end{picture}}
\caption{({\sl left}) ACIS-S X-ray intensity along the same major axis slice as in Fig~\ref{ima-prof-HRC-I.00}. 
({\sl right}) Index spatial profile along the same slice. Coordinate origins of the spatial axes coincide 
with the pulsar position, and the vertical dotted line marks the position of the blob. Southwest is to the right of the
plots. The intensity and index maps were smoothed with a one ACIS pixel Gaussian kernel before the sampling along 
the slice was made.
}
 \label{X-ray_index}
 \end{figure*}

\begin{figure*}
\begin{center}
\includegraphics[height=180mm,angle=270, clip]{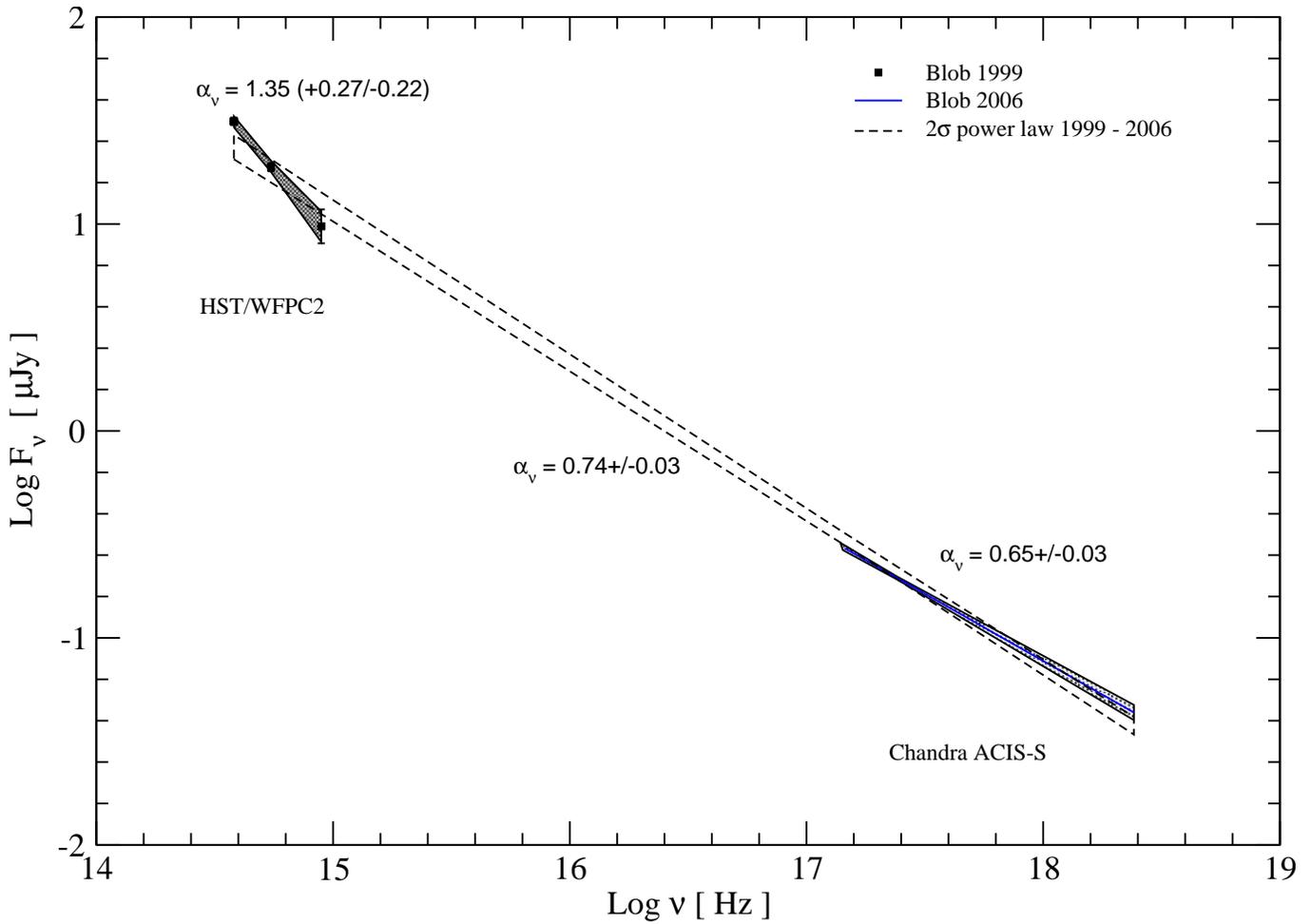}
\end{center}
\caption{Multiwavelength spectrum for the ``blob'' region. See text for details.
}\label{f:MWblobs}
\end{figure*}
 As seen in Figure~\ref{X-ray_index}, the blob demonstrates the hardest emission within
the PWN. Its hardness even exceeds that of the pulsar, underlining the peculiarity of
this structure. In general, the brighter the region, the harder its spectrum. This leads to 
the left-right asymmetry of the hardness of the PWN along the major axis clearly seen in
Figure~\ref{X-ray_index}, where the orientation of the  NE-SW major axis slice is the same way 
as in Figures~\ref{polSlice:2007},~\ref{ima-prof-547M.99}, and~\ref{ima-prof-HRC-I.00}.
The hardness profile along the minor PWN axis is, on the other hand, almost symmetrical,
 and demonstrates a gradual softening of the emission with distance from the pulsar, from 
$\alpha$$\sim$0.9 to $\alpha \gsim 1.6$. 
The hardness image (Fig.~\ref{f:indexGammaIma1}) does not reveal any hardening to extreme values for 
the assumed torus region, as seen in the Crab PWN \citep{Weisskopf00,Mori04}. 

\subsection{Multiwavelength spectrum of the ``blob" region}
In Serafimovich et al. (2004) we studied the optical spectra for the 1999 HST/WFPC2 
epoch of several positions in what we then defined the torus and jet of the PWN. 
One of these positions, called ``Area 2" in Serafimovich
et al. (2004), turns out to partly overlap with the ``blob'' discussed by De Luca et al. (2007).
The results of Serafimovich et al. (2004) hint that the optical spectrum of their ``Area 2" in 1999 
was different from other regions of the PWN since it could have had a somewhat steeper 
spectrum in the optical with $\alpha_{\nu}=1.58^{+0.33}_{-0.30}$. In Table 4, the 
spectral slope of the ``blob" has $\alpha_{\nu}=1.35^{+0.27}_{-0.22}$ for the same epoch. 
The overlap between ``Area 2" in Serafimovich et al. (2004) and the ``blob" in Table 4 is
only $\sim 60$\%. The results are, however, clearly compatible even within 1$\sigma$.
How the optical emission of the ``blob" varies with time, if moving, as well as other 
regions how they vary with time, is discussed in Sect. 3.2, and listed in Table~\ref{optical:fits}.

Perhaps of more general interest is to see how the optical emission connects to the
emission at other wavelengths. In Serafimovich et al. (2004, their Fig. 15), a comparison
between the multiwavelength spectrum (radio to X-rays) from the entire PWNe of the Crab and 
0540 was made. For both objects, an extension of the X-ray spectrum overshoots the 
optical spectrum in a log($\nu$) vs. log($F_{\nu}$) diagram. A further roll-off towards the radio
is seen in both PWNe. A cubic spline fit for the entire logarithmic spectrum for the Crab PWN, 
overshoots both in the optical and radio for 0540, if fit to its X-ray part.
We have done the same multiwavelength study here, but concentrating on the 0540 ``blob". 
The optical data are again from 1999, but for the X-ray data, we used our 
reductions of the 2006 Chandra data. These have better signal-to-noise than earlier 
data. We could have chosen optical data closer in time to the 2006 Chandra data, but 
a look at Table~\ref{optical:fits} for all areas, except the ``blob", shows that the optical 
flux is stable between 1999 and 2007. For the ``blob" in Table~\ref{optical:fits} we applied 
$0\farcs4$ shift between 1999 and 2007, so the overlap between the $0\farcs455$ apertures
for each epoch is fractional. Nevertheless, the flux in the red HST bands for the ``blob"
in Table~\ref{optical:fits} are similar between 1999 and 2007 to within $5-10$\%. 
We emphasize that the X-ray spectrum of the blob has been extracted from the same 
aperture position and extent as has been used for the 1999 optical flux measurements. 
For the absorbing gas column density $N_{\rm H}$ we used the results of 
Sect. 3.5. 

The multiwavelength result is shown in Fig.~\ref{f:MWblobs}. It is clear that an 
extension of the X-ray data, considering the rather well-determined X-ray slope, 
undershoots rather than overshoots the optical data. What stands out here, like in 
Serafimovich et al. (2004) for the whole PWN, is the intrinsically steep optical spectrum 
compared to the X-ray spectrum. We caution that the errors calculated, and shown in 
Fig.~\ref{f:MWblobs} for the optical data points, are the statistical errors only. If we
would allow for for a modest ($12-15$ \%) systematic error of the F791W flux,
one can connect all optical and X-ray data with a single power-law to within 2$\sigma$.
The slope of that power-law would be $\alpha_{\nu}=0.74\pm0.03$, which is
marginally different from the X-ray spectral slope with $\alpha_{\nu}=0.65\pm0.03$. 
As we noted above, at least part of the systematic error ($5-10$\%) may come from using 
the spatially slightly shifted 1999 optical data, rather than data closer to the epoch of the 
X-ray data, i.e., 2006. 

 We emphasize that our finding that a single power-law may fit the optical/X-ray part
of the spectrum for 0540 was derived only for the ``blob", and therefore does not
necessarily contradict our findings in Serafimovich et al. (2004) that spectral breaks are
needed for the whole PWN. While we leave a discussion for the entire nebula for a future 
paper, we can make use of Figs.~\ref{ima-prof-547M.99},~\ref{ima-prof-HRC-I.00} 
and~\ref{f:indexGammaIma1} 
to check whether our reductions of Chandra data make a difference compared
to the X-ray spectra used in Serafimovich et al. (2004). From Fig.~\ref{f:indexGammaIma1}
we note that the X-ray spectral index is in the range $1.4-1.8$ in the north-west and south-east
part of the PWN, and from Figs.~\ref{ima-prof-547M.99} and~\ref{ima-prof-HRC-I.00} that the
optical and X-ray fluxes are $\sim 2$ and $\sim 3-5$ times weaker there, respectively, than at 
the ``blob" position. In Fig.~\ref{f:MWblobs} we show that a single power-law with spectral index 
$\alpha_{\nu}=0.74\pm0.03$ could be enough for the ``blob". Using a difference of 3 dex
between the optical and X-ray frequencies, an X-ray spectral index of $\alpha_{\nu}=1.6$
would make an extension of the X-ray spectrum overshoot the optical by more than 2 dex
in flux in the north-west or south-east parts of the PWN. A single power-law for those regions
is thus not possible, or actually in any part of the remnant where the X-ray spectral index is 
$\alpha_{\nu} \gsim 0.9$. As $\alpha_{\nu} \gsim 0.9$ is true for most of the PWN, a single
power-law for the whole PWN will not be possible, and our results will in this sense not
contradict the findings of Serafimovich et al. (2004). From Fig.~\ref{f:indexGammaIma1}
we note that $\alpha_{\nu} \lsim 0.9$ applies only along the major axis within $\sim 2$ 
arcseconds from the pulsar. The simplest explanation to this is that energy injections
occur in this region, and that cooling of electrons take place, thus steepening the X-ray 
spectrum, when electrons move away from this region.

\section{Discussion}
\subsection{Morphology of the PWN}\label{morph}

The PWN of \psr\ and \snr\ have been imaged on several occasions. 
In \citet{Seraf04} we concentrated on the continuum-emitting PWN, and 
images showing the line-emitting gas of the inner $\sim 4$\asec\ of the SNR are nicely
compiled in \citet{Morse06}. While there are some general features in
common between continuum-emitting and line-emitting gas (like overall shape
and concentration toward the southwest), there are also several important
differences. 

\begin{figure}
\begin{center}
\includegraphics[width=80mm, clip]{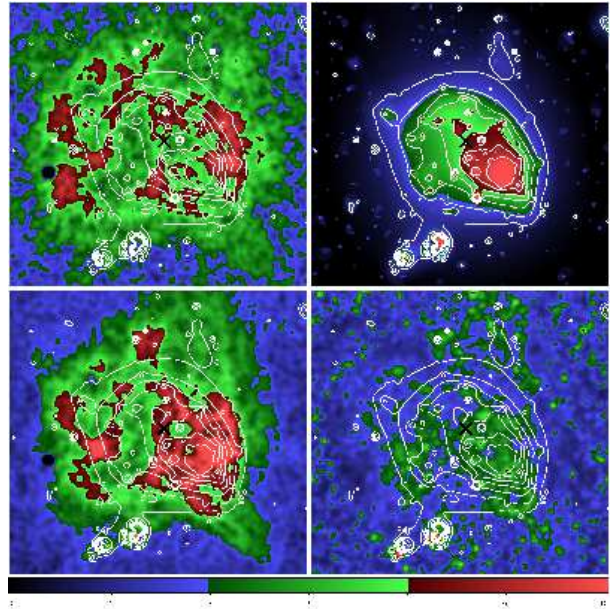}
\end{center}
\caption{{\it Upper left:} Wavelet filtered [O~III] image of PWN 0540 
obtained with HST/F502N in 1999 with overlaid contours from the continuum 
image in the upper right panel. {\it Upper right:} Wavelet filtered continuum 
image of PWN 0540 obtained with HST/F547M in 1999 with overlaid contours. 
The black cross shows the pulsar position. {\it Lower left:} Wavelet 
filtered [S~II] image obtained with HST/F673N in 1999 with overlaid 
contours from the continuum image. {\it Lower right:} Ratio map of  [S~II] / [O~III], 
with [S~II]  and [O~III] from the lower left and upper left panels, respectively. 
Overlaid contours are from the optical continuum in the upper right panel.
The scale at the bottom of the figure represents color coding for that ratio image.
}\label{f:ContIma1}
\end{figure}

The line-emitting gas is concentrated to the 
south-west, but with distinctions between the spatial distribution of, 
e.g., [O~III] and [S~II]. This can be seen in the wavelet filtered images from 1999 
shown in Fig. \ref{f:ContIma1} . The [O~III] structure appears overall rounder in shape, 
whereas [S~II] is more focussed to the southwestern conglomerate, and shows 
only little emission to the northeast. A protrusion extending  $\sim 2$\asec\  
outside the main body of the emission in the southwestern direction is seen
in both [O~III] and [S~II]. The structure of this is reminiscent of the Crab chimney.
What is striking is that the protrusion connects to the area of strongest [S~II]
emission in the remnant body, whereas the [O~III] emission in the remnant is in
general weaker in the pulsar-protrusion direction, than in the areas east and west of 
it, i.e., the areas flanking the ``blob". 

The optical continuum emission from the same epoch, as exemplified by the 
uppermost right panel in Fig. \ref{f:ContIma1} and contours
overlaid on the line-emission maps, is even more concentrated to the southwest
than the [S~II] emission, but the overlap between [S~II] and the continuum is obvious. 
No clear indication of a protrusion is seen in the wavelet filtered continuum image, which 
could be due to lower signal-to-noise. The overlap between optical continuum and regions of 
enhanced [S~II] emission is evidenced by the lower right panel of Fig. \ref{f:ContIma1}, 
where we have made a ratio map of [S~II]  and [O~III]. The highest optical continuum flux 
occurs in a conglomerate with a large [S~II] / [O~III] ratio. The correlation between optical 
continuum and the [S~II] / [O~III] ratio is indeed even more pronounced than with the [S~II] 
emission alone. This could either be a physical 
coincidence of continuum emission and regions of high sulphur abundance and/or
lower state of ionization, or it could signal that shock excitation is more important
in those regions compared to regions with smaller [S~II] / [O~III] ratios. Shocked
gas generally emits more [S~II]  than [O~III] compared to photoionized gas, and the
temperature of the gas is higher (see Williams et al. 2008). A temperature higher than
the equilibrium temperature in photoionization dominated plasma can also be
obtained in a time dependent situation. An example is the flash photoionization of the
inner ring around SN 1987A where the [O~III] temperature was $\sim 5\times10^4$~K
shortly after the outburst (see Fig. 1 in Lundqvist \& Fransson 1996). We explore
this possibility further below.

\begin{figure}
\begin{center}
\includegraphics[width=80mm, clip]{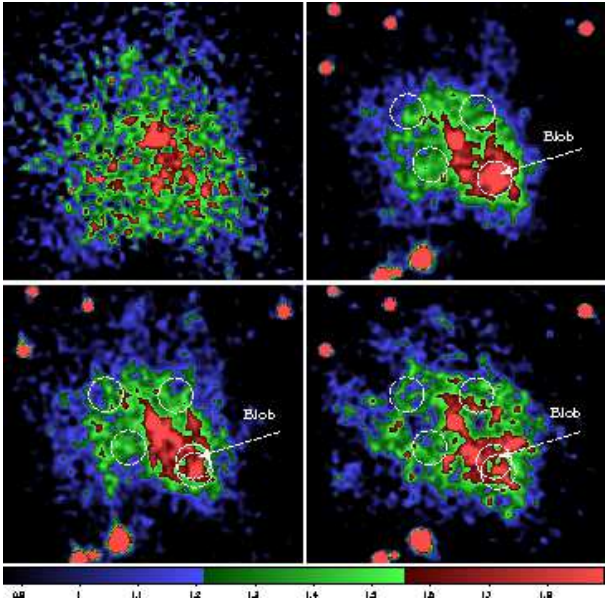}
\end{center}
\caption{The \psr\ and its PWN obtained with HST/F547M in 1992 ({\it upper 
left}), in 1999 ({\it upper right}), in 2005 ({\it lower left}) and HST/F450W 
in 2007 ({\it lower right}). Note that the cut levels for the F540W image were 
compensated for the slope of the nebula's SED. The white circular apertures show
the ``blob'' position in 1999 and respective epoch, as well as three other areas 
selected for comparison. In the 2005 and 2007 images we have also made circular
apertures for the 2005 position of the ``blob'', and the expected 2007 position if the
``blob'' displacement would be linear in time. Some stars projected on the PWN were 
subtracted off. 
}\label{f:CompIma1}
\end{figure}

\begin{figure}
\begin{center}
\includegraphics[width=80mm, clip]{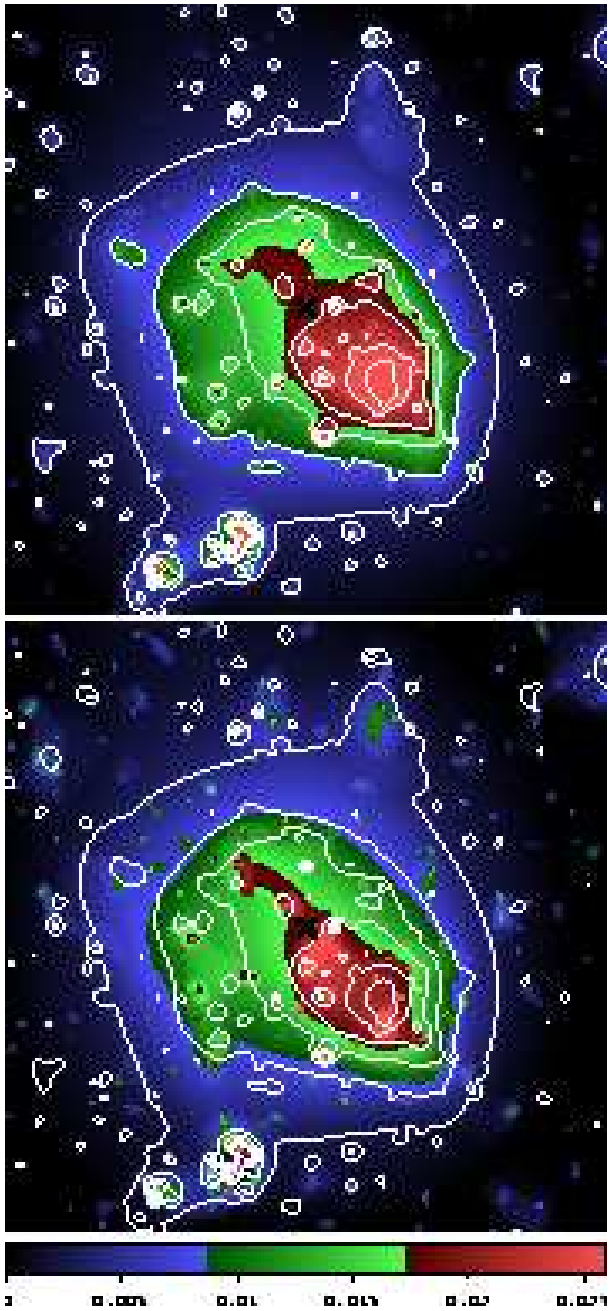}
\end{center}
\caption{{\it Upper:} Wavelet filtered continuum image of PWN 0540 obtained 
with HST/F547M in 1999 with overlaid contours. {\it Lower:} Wavelet filtered 
continuum F547M in 2005 image with overlaid contours from the 1999 epoch.
 Note how the ``blob" structure and the structure just southwest of the pulsar change 
in morphology between these two epochs. For the structure close to the pulsar the strongest 
emission in the 2005 image is much wider than in the upper panel. The increased emission
close to the pulsar could be new blob emerging.
}\label{f:CompContwav}
\end{figure}

The spatial distribution of the continuum emission in the PWN of 0540 shows high 
variability in the optical. This was noticed by De Luca et al. (2007), and they argued that 
the ``blob'' shown in Fig. \ref{f:CompIma1} had moved between 1999 and 2005. The change 
in the spatial distribution of the optical continuum is clearly seen in our wavelet filtered images in 
Fig.~\ref{f:CompContwav}. Figure  \ref{f:CompContwav} also shows how a new structure
appears just southwest of the pulsar, as if a new ``blob'' would be generated, whereas the
overall structure does not differ between the two epochs. In addition, Fig. \ref{f:CompContwav} 
shows how the structure extending from the pulsar (marked by a black cross) toward the ``blob'' 
continues in the northeastern direction, although with lower intensity and less extent.

The soft X-ray emission has an even more remarkable concentration of emission
to the ``blob" position than in the optical, with a soft X-ray flux as bright as $\sim 10$\% of the 
pulsar emission. The X-ray position does not seem to vary appreciably with time between 2000 and 2006, 
and the X-ray spectrum is a power-law which is hardest at the ``blob" position, and becomes
increasingly softer away from it (see Fig.~\ref{f:indexGammaIma1}). In general, the spectrum is harder 
along the NE-SW direction than off this orientation. The optical continuum emission shares the general 
brightness distribution of the X-ray emission along the NE-SW direction, but it does also show 
strong variability of the spatial distribution. That we are able to see this in the optical, but not in X-rays, 
may be an effect of better spatial resolution. 

\subsection{Excitation of the line emitting filaments close to the ``blob"}\label{entity}
The variability of the spatial distribution of the optical {\it line emission} is less known since 
observations through HST line filters have only been performed once, namely in 1999 
(Morse et al. 2006). However, we also have some slit spectra taken in 1996, which partly sample
the ``blob" region (N. Lundqvist et al., in preparation), and which do not indicate strong [S~II] emission 
at the position of the 1999 ``blob''. This, together with the fact that the optical continuum emission 
does vary, suggests that the line emission may also vary with time. 

The excitation of the line emitting filaments is most naturally explained 
through shock activity (Williams et al. 2008), but in Sect. 4.1 we also highlighted the possibility 
of time dependent photoionization. The idea is that the optical continuum activity
could also indicate higher X-ray activity, and that the X-rays could photoionize unshocked 
filaments that are quasi-stationary.
As we have seen, the X-ray emission from the ``blob'' in 2006 is about a factor of 10 weaker 
than the emission from the pulsar, but we only wish to study the photoionization close to
the ``blob'' where X-rays from the ``blob'' dominate, so to study whether time-dependent 
photoionization is important, we have ignored the pulsar X-rays. 

To test the effect of time dependence we have made models where we have
artificially switched on the X-ray emission from the ``blob'' to study how it would 
photoionize nearby gas. The model is similar to that used in Lundqvist \& Fransson 
(1996) for SN 1987A, but with several updates (see e.g. Mattila et al. 2010). We 
have assumed that the ``blob'' is spherical with radius 0\farcs2, and that the gas being 
photoionized surrounds the ``blob'' also in a spherical fashion. For the density of this gas 
we adopted the results of Serafimovich et al. (2005) who derived an electron density 
in the range $1-5 \times 10^3$ cm$^{-3}$ from the [S~II]~$\lambda\lambda$ 6716-31
lines. For the abundances of the filaments we have assumed a composition with
the relative abundances (by number) suggested by Williams et al. (2008), i.e., 
C: O: Ne: Mg: Si: S: Ar: Ca: Fe = 0.1: 1: 0.2: 0.1: 0.1: 0.1: 0.1: 0.1: 0.1. We have
also blended in H and He at various amounts as Serafimovich et al. (2005, see
also Morse et al. 2006) clearly revealed the presence of hydrogen. 
The ionizing luminosity from the ``blob'', when switched on, is assumed to 
be $4.66 \times 10^{31} (E/1~{\rm keV})^{-0.65}$ erg s$^{-1}$ eV$^{-1}$.
With H and He mixed in so that H: He: O = 1: 1: 1, and with a total density of
atoms and ions $n_{\rm at} = 1.5 \times 10^3$ cm$^{-3}$, the ionization parameter
(number density of ionizing photons divided by $n_{\rm at})$ is $\sim 10^{-3}$, i.e.,
only a thin rim of gas is photoionized around the ``blob''. Steady state is reached in
$\sim 10$ years, which is longer than the time scale for changes in the ``blob'' structure,
justifying time dependence. Peak temperature of the ionized gas reaches 
$\sim 8.5\times10^3$ K, and occurs already after 3 months. Maximum ionization
occurs when steady state is being approached. Although sulphur is mainly in S~II, i.e.,
which we know from Figure~\ref{f:ContIma1} is just what is observed in the ``blob" region,
most of the oxygen is still in O~I. This contrasts the observations of Kirshner et al. (1989) 
which show very weak [O~I] emission. One must caution that the observations
of Kirshner et al. were made more than one decade before the ``blob" in 1999, and that the
slit of Kirshner et al. was not ideally placed to probe the ``blob" region. The very small ionization 
parameter and the low temperature of the ionized gas are, however, enough to make us
rule out photoionization by X-rays from the ``blob'' structure as a viable source for
the excitation of the [S~II] emitting gas in the southwestern part of the PWN. If we
disregard the possibility that this region is truly more abundant in sulphur than
other parts of the nebula, the remaining explanation is what is discussed by
Williams et al. (2008), i.e., [S~II] being boosted by shock activity.

\subsection{Is the ``blob" a moving entity?}\label{entity}
In Fig. \ref{f:CompIma1} we also include the epochs 1992 and 2007. The signal-to-noise is 
too low in the 1992 data, which we have included for completeness, to draw any conclusions 
more than that the continuum emission is enhanced in the southwestern direction. The 2007
image shows a much weaker ``blob'', and it is in fact not obvious that the ``blob'' structure is longer 
present in 2007.  In general, the U-shaped structure in the 1999 image, with the pulsar 
at the upper left part of the U-structure, and the ``blob'' at the bottom, remains roughly intact between
1999 and 2005, but breaks up in 2007. A ``blob''-structure so clearly seen in 2005 divides
up into three fragments with the lower one roughly at the expected position of the ``blob'', but with
the strongest emission seen northeast of the 1999 ``blob'' position. In 2005 that position is actually
devoid of strong emission. One could imagine that this new structure in 2007 could signal
propagation of the 2005 ``blob'' just southwest of the pulsar, but we also note that the structure
just northeast of the ``blob'' in 2007 is clearly seen in 1999, but less brighter. A possibility is that
the whole structure responsible for the continuum emission exists at all epochs, but that it
contributes with different amounts at different epochs. With this interpretation, the northern
part of the ``blob'' is just an enhancement in emission in 1999, which then fades away, leaving
only the southern part of the ``blob'' seen in 2005, and even more so in 2007.

Our findings can therefore suggest that the ``blob'' structure may not be a moving entity, 
but rather a part of the filamentary structure that was excited in 1999, and then fades in 
optical intensity. Particularly evident is how well the continuum emission traces the 
emission structure in [S~II] and how low the correlation is between continuum and 
[O~III] (see Sect. 4.1). The high [S~II]/[O~III]-ratio suggests shocks (cf. Sect. 4.2), 
this possibility being strengthened by the fact that the ``blob'' coincides with 
a position in the PWN where the polarization angle changes abruptly and where there 
is strong X-ray emission. A possibility is an outflow in the PWN from the pulsar which 
excites the filaments at the ``blob'' position in a time-varying fashion. Another possibility
is a local release of energy at the ``blob'' position which then causes a point-like 
expansion of plasma. In both these cases there will be, after some delay, slow shocks 
transmitted into filaments embedded in the PWN in a way similar to the shocks transmitted 
into the blobs of the circumstellar ring around SN 1987A (e.g., Gr\"oningsson et al. 2008).

As discussed in Sect. 3.3, there are several positions along the major symmetry axis
(NE-SW) of the PWN where the polarized flux peaks other than at the ``blob" position, and where
the polarization angle changes at these positions. This could be signs of past, or weak
present, shock activity. A possibility is that energy is released along this symmetry axis
at different places at different epochs, which means that the ``blob" was just a recent
such event. If the ``blob''-like feature seen in 2005 just southwest of the pulsar 
is a new feature moving southward, the sporadic energy release scenario may prove 
wrong. If it breaks up, it could speak in favor of a sporadic energy release. Indeed, we note in 2007
a peak above 3$\sigma$ at $\sim 0\farcs6$ in polarized flux, southward along the
major symmetry axis (see Fig. 8), as well as a change in polarization angle. Looking at Fig.~\ref{f:CompIma1}
from 2007, no obvious blob-like counterpart is seen at that position. This could hint
that the 2005 ``blob''-like feature broke up, leaving only an imprint on the polarization map.

We also note that the two strong peaks in polarized flux southwest of the pulsar occur in the
protrusion-like feature most clearly seen in lower left panel of Fig.~\ref{f:ContIma1} which shows
wavelet-filtered [S~II] emission from 1999. The NE-SW slice in Fig.~\ref{ima-prof-547M.99} catches 
the northern part of this protrusion.
Further polarization studies would shed more light on this and other regions of the PWN. 
At the moment we only have good data from one epoch, namely the one from 
2007 discussed here. 

The strong spatial variation of the inferred X-ray power-law slope of the emitted radiation 
(cf. Fig.~\ref{f:indexGammaIma1}), shows that multiwavelength spectra should be studied over 
limited spatial extents. In Serafimovich et al. (2004) we invoked possible spectral breaks between 
the optical and X-rays when studying the multiwavelength spectrum of the full PWN. 
From Fig.~\ref{f:MWblobs} here, we argue that a single power-law may be sufficient for the ``blob".
A single power-law could mean that the 1999 ``blob" position is a site of
energy injection to relativistic electrons. Further multiwavelength, and preferably contemporaneous, 
studies at other positions, away from the ``blob", would reveal if a break in the power-law appears,
and if cooling of the electrons occurs. We leave this for a future study.

\subsection{The supernova progenitor}\label{progenitor}
The X-ray spectral analysis  presented in Sect. 3.5 gave better results if we included
``extra" absorption from oxygen-rich gas in addition to the interstellar gas in the LMC and
Milky Way. The column density of extra oxygen we found to be needed was 
$N^{SNR}_{\rm O} \approx 3.1\times$10$^{18}$ cm$^{-2}$. Adopting SN 1987A, and particularly 
the model used in Blinnikov et al. (2000) and Serafimovich et al. (2004) as a model for 0540, 
$N^{SNR}_{\rm O} =9.5\times$10$^{17}$ cm$^{-2}$ for an SN 1987A age of $t=800$ years, where
the column density scales with age as $t^{-2}$. It seems unlikely that 0540 is younger than 
800 years, so if $N^{SNR}_{\rm O}$ is indeed as high as suggested from our X-ray analysis,
then more oxygen was produced in 0540 than in SN 1987A, and/or the oxygen-rich ejecta of 0540 
expand more slowly or it could be differently mixed than in the model of Blinnikov et al. (2000).
As we do not find any evidence for strong variation of the oxygen column density over the PWN, 
a higher than average column density along the line of sight to us is less likely.

It is interesting to note that the model used in Blinnikov et al. (2000) invokes $\sim 2 \Msun$ of oxygen,
whereas Williams (2010) in his PhD thesis argues for an upper limit of $\sim 3.5 \Msun$ of oxygen.
If the oxygen mass is this high in 0540, and the velocity structure of oxygen more ``compact" than 
in SN 1987A, $N^{SNR}_{\rm O}$ could be close to the value inferred from our
X-ray extractions. A high oxygen mass would then favor a progenitor in the upper mass range
for the ones discussed by Chevalier (2006), and we would end up with the same mass range of
$20-25 \Msun$ as favored also by Williams (2010). We note that Park et al. (2009) did not 
need to include elevated levels of oxygen for the X-ray absorption, which does not contradict
our indications. Their analysis was made for the outer parts of the supernova remnant not
covered by the oxygen-rich ejecta. 

\subsection{Future observations}\label{future}
To map out the 3D shock structure of the filaments, good resolution both in space and
velocity is needed. Previous deep spectroscopic observations of 0540, partly used in 
Serafimovich et al. (2004, 2005) and discussed in a forthcoming paper (N. Lundqvist et al.,
in preparation, see also above), show no clear [S~II] activity close to the ``blob'' position, but 
these observations were made prior to the knowledge about the ``blob'', and the slits were not 
ideally placed to constrain the ``blob'' activity. These observations were also carried out at 
seeing $\gsim 1\farcs1$. To map out the velocity structure of the ``blob'' region in order to 
constrain the propagation of the shocks, observations with good spatial resolution and coverage 
are needed. Modern Integral-field-units (IFUs) with a spectral resolution of ($\lsim 100 \kms$)
are particularly useful, especially in good seeing and at low airmass. The 
good velocity resolution is useful to separate out slow (tens of $\kms$) shocks driven 
into the filaments from the faster ($\sim 250\kms$) shocks driven into more hydrogen-rich gas 
(cf. Williams et al. 2008). Observations in the optical should also be sensitive enough
to catch weaker lines like [O~III]~$\lambda$4363 for temperature estimates.

The ``blob'' in the southwest part of the PWN, first seen in 1999, was truly an extraordinary 
event. As we have argued for above, similar events may, however, have occurred also in the 
past. Continuous imaging monitoring of 0540 is therefore important, both in the continuum and 
in lines like [S~II] and [O~III], to see if new blobs emerge. HST can continue to play an important 
role here, especially with the upgrade to WFC3 which has no CTE effects. To see how polarization 
evolves would be particularly useful for a global picture of the PWN. Further down the line, both 
JWST and ALMA with their good sensitivity and spatial resolution will provide very important links 
between optical and radio studies of the PWN of 0540. Such studies at good angular resolution
will also make it more easy to compare 0540 to its much closer twin, the Crab.\\

Acknowledgements $-$ 
 We would like to thank the referee for important remarks that helped to clarify the text. We also
thank Jennifer Mack Research and Instrument Scientist in the ACS/WFPC2 team at the Space 
Telescope Science Institute for the assistance with calibration of the HST/WFPC2 data, and 
Jean-Luc Starck for discussions. PL and CIB acknowledge support from the Swedish Research 
Council, and GO support from Swedish National Space Board. YS and DZ were partly supported 
by RFBR (grants 08 02 00837 and 09 02 12080) and by the State Programm  "Leading Scentific 
Shcools of RF" (grant NSh 3769 2010.2). DZ was also supported by a St. Petersburg Government 
grant for young scientists.

{}
  
\label{lastpage}
\end{document}

%% file: tab21.tex
\begin{table*}
\caption{HST/WFPC2 optical observations of the \psr\ field used in our study.}
\label{t21:arch} 
\begin{center}
\begin{tabular}{llccrcl}
\hline
\hline
Date & Total &\multicolumn{3}{c}{Filter} & Proposal & Comments\\ 
     & exp.$^a$ &    name & PHOTPLAM$^b$ & PHOTBW$^c$ & ID & \\
\hline
17Nov92$^d$& 2x400 & F547M & 5483.9 & 205.5 & 4244 & continuum\\
\hline
19Oct95 & 2x300  & F555W & 5442.9 & 522.2 & 6120 & overlaps with [O~III]$\wl$ 5007\\
\hline
17Oct99 & 2x300  & F336W & 3359.5 & 204.5 & 7340 & continuum\\
17Oct99 & 8x1300 & F502N & 5013.3 &  48.3 & $\cdots$& centered on [O~III]$\wl$ 5007\\
17Oct99 & 2x400  & F547M & 5483.9 & 205.5& $\cdots$ & continuum\\
17Oct99 & 6x1300 & F673N & 6732.3 &  30.7& $\cdots$ & centered on [S~II]$\wll$ 6716,6731\\
17Oct99 & 2x200  & F791W & 7872.5 & 519.8& $\cdots$ & continuum\\
\hline
15Nov05$^e$ & 4x260 & F547M & 5483.9 & 205.5& 10601 & continuum\\
15Nov05$^e$ & 3x160 & F555W & 5442.9 & 522.2 & $\cdots$ & overlaps with [O~III]$\wl$ 5007\\
\hline
21Jun07 & 3x260 & F336W & 3359.5 & 204.5 & 10900 & continuum\\
21Jun07 & 3x260 & F450W & 4557.3 & 404.2& $\cdots$ & continuum\\
21Jun07 & 3x100 & F555W & 5442.9 & 522.2& $\cdots$ & overlaps with [O~III]$\wl$ 5007\\
21Jun07 & 3x140 & F675W & 6717.7 & 368.3& $\cdots$ & overlaps with [S~II]$\wll$ 6716,6731\\
21Jun07 & 3x200 & F814W & 7995.9 & 646.1& $\cdots$ & continuum\\
\hline
\end{tabular} \\
\begin{tabular}{ll}
$^a$ \  Total exposure is given in number of images times the & $^b$ \  Pivot wavelength of the filter band measured in \AA. \\
~~~~exposure time of individual images in seconds. & $^c$ \  Width of the filter band measured in \AA. \\
$^d$ \ The 1992 observations were obtained with HST/WFPC & $^e$ \ Our observations. \\
~~~~instead of HST/WFPC2. \\
\end{tabular}
\end{center}
\end{table*}

%% file: tab22.tex
\begin{table}
\caption{Chandra X-ray observations of the \psr\ field used in our study.}
\label{t3:arch} 
\begin{center}
\begin{tabular}{cccc}
\hline
\hline
Date & Exposure & Instrument & ObsID \\
 & (ks) & & \\
\hline
31Aug99 & 17.79 & HRC-I & 132 \\
\hline
21Jun00 & 10.04  & HRC-I & 1735 \\
21Jun00 & 10.04  & HRC-I & 1736 \\
22Jun00 & 10.07  & HRC-I & 1741 \\
\hline
15Feb06 & 40.36  & ACIS-S & 5549 \\
16Feb06 & 39.98  & ACIS-S & 7270 \\
18Feb06 & 38.56  & ACIS-S & 7271 \\
\hline
\end{tabular} 
\end{center}
\end{table}

%% file: tab23.tex
\begin{table}
\caption{HST/WFPC2 polarization observations of the \psr\ field used in our study.}
\label{t23:arch} 
\begin{center}
\begin{tabular}{cccrc}
\hline
\hline
Date & Total & Mode of obs. & Pol. & Proposal \\
 & exp.$^a$ & & angle & ID \\
\hline
05Nov07 & 600x3 & F606W$+$POLQ &  0\grad & 10900 \\
25Sep07 & 600x3 & F606W$+$POLQ & 45\grad & 10900 \\
26Jul07 & 600x3 & F606W$+$POLQ & 90\grad & 10900 \\
21Jun07 & 600x3 & F606W$+$POLQ &135\grad & 10900 \\
\hline
\end{tabular} 
\begin{tabular}{l}
$^a$ \  The total exposure is given in number of images times \\ 
~~~~the exposure time of individual images in seconds. \\
\end{tabular}
\end{center}
\end{table}

%% file: table-optf.tex
\begin{table*}
\begin{center}
\caption{Results of optical spectral  fits for several regions of the 0540 
pulsar+PWN system. Columns 6 and 7 are the measured and dereddened fluxes 
with E(B-V)=0.20 (AV=0.62), respectively.  } 
\label{optical:fits}
\begin{tabular}{cccclllcc}
\hline 
\hline
Epoch & \multicolumn{2}{c}{Source} & \multicolumn{2}{c}{Band} & \multicolumn{2}{c}{log(Flux)} & \multicolumn{2}{c}{power law} \\
 & Region  & Area & name & Log($\nu$) & observed & dereddened & $\alpha_\nu$$^a$ & norm. const.$^b$ \\
 & & arcsec$^2$ & & Hz & $\mu$Jy & $\mu$Jy & & $\mu$Jy \\
\hline \\[-2mm]
1999 & Blob & 14\farcs3 & F791W & 14.581(15) & 1.343(18) & 1.496(18) & 1.35$(^{+27}_{-22})$ & \\
$\cdots $ &$\cdots $ & $\cdots $& F547M & 14.738(8) & 1.026(20) & 1.275(20) & $\cdots $ & 1.28$(^{+3}_{-3})^c$ \\
$\cdots $ & $\cdots $& $\cdots $& F336W & 14.950(13)& 0.584(82) & 0.989(82) & $\cdots $ & \\[+2mm]
2007 & Blob & 14\farcs3 & F814W & 14.574(18) & 1.255(20) & 1.403(20)& 1.06$(^{+18}_{-16})^c$ & \\
$\cdots $ & $\cdots $& $\cdots $& F675W & 14.649(12) & 1.160(22)& 1.357(22) & $\cdots $ & \\
$\cdots $ & $\cdots $& $\cdots $& F555W & 14.741(21) & 0.974(26)& 1.224(26) & $\cdots $ & 1.24$(^{+2}_{-2})$ \\
$\cdots $ & $\cdots $& $\cdots $& F450W & 14.818(20) & 0.837(26)& 1.152(26) & $\cdots $ & \\
$\cdots $ & $\cdots $& $\cdots $& F336W & 14.950(13) &-0.006 & 0.398$^d$  & & \\
$\cdots $ & $\cdots $& $\cdots $& F336W & 14.950(13) & 0.188(72)& 0.593(72)$^e$  & & \\
$\cdots $ & $\cdots $& $\cdots $& F336W & 14.950(13) & 0.510(50)& 0.915(50)$^f$  & & \\[+2mm]
\hline\\[-2mm]
1999 & Anti-blob & 14\farcs3 & F791W & 14.581(15) & 1.012(28) & 1.164(28)& 0.90$(^{+35}_{-27})$ & \\
$\cdots $ &$\cdots $ & $\cdots $& F547M & 14.738(8) & 0.707(32)& 0.955(32) & $\cdots $ & 1.00$(^{+3}_{-4})^c$ \\
$\cdots $ & $\cdots $& $\cdots $& F336W & 14.950(13)& 0.408(108)& 0.813(108) & $\cdots $ & \\[+2mm]
2007 & Anti-blob & 14\farcs3 & F814W & 14.574(18) & 1.036(26) & 1.184(26) & 1.00$(^{+33}_{-25})$ & \\
$\cdots $ & $\cdots $& $\cdots $& F675W & 14.649(12) & 0.976(26)& 1.173(26)&  & \\
$\cdots $ & $\cdots $& $\cdots $& F555W & 14.741(21) & 0.742(38)& 0.992(38)&  & \\
$\cdots $ & $\cdots $& $\cdots $& F450W & 14.818(20) & 0.641(36)& 0.956(36)& $\cdots $ & 0.94$(^{+5}_{-6})$ \\
$\cdots $ & $\cdots $& $\cdots $& F336W & 14.950(13) & 0.382(118)& 0.787(118)& $\cdots $ & \\[+2mm]
\hline\\[-2mm]
1999 & Area 3 & 14\farcs3 & F791W & 14.581(15) & 1.071(32) & 1.224(32)& 1.30$(^{+44}_{-33})$ & \\
$\cdots $ &$\cdots $ & $\cdots $& F547M & 14.738(8) & 0.763(34) & 1.012(34) & $\cdots $ & 1.02$(^{+4}_{-5})^c$ \\
$\cdots $ & $\cdots $& $\cdots $& F336W & 14.950(13)& 0.318(132)& 0.723(132)& $\cdots $ & \\[+2mm]
2007 & Area 3 & 14\farcs3 & F814W & 14.574(18) & 1.086(28) & 1.234(28)& 1.32$(^{+42}_{-31})$ & \\
$\cdots $ & $\cdots $& $\cdots $& F675W & 14.649(12) & 1.035(28)& 1.232(28)&  & \\
$\cdots $ & $\cdots $& $\cdots $& F555W & 14.741(21) & 0.799(36)& 1.049(36)&  & \\
$\cdots $ & $\cdots $& $\cdots $& F450W & 14.818(20) & 0.753(30)& 1.067(30)& $\cdots $ & 0.95$(^{+5}_{-8})$ \\
$\cdots $ & $\cdots $& $\cdots $& F336W & 14.950(13) & 0.276(146)& 0.681(146)& $\cdots $ & \\[+2mm]
\hline\\[-2mm]
1999 & Area 4 & 14\farcs3 & F791W & 14.581(15) & 1.074(26) & 1.226(26)& 1.00$(^{+35}_{-28})$& \\
$\cdots $ &$\cdots $ & $\cdots $& F547M & 14.738(8) & 0.796(28) & 1.045(28)& $\cdots $ & 1.06$(^{+3}_{-3})$ \\
$\cdots $ & $\cdots $& $\cdots $& F336W & 14.950(13)& 0.436(108)& 0.841(108)& $\cdots $ & \\[+2mm]
2007 & Area 4 & 14\farcs3 & F814W & 14.574(18) & 1.098(24)& 1.246(24)& 0.86$(^{+26}_{-21})$& \\
$\cdots $ & $\cdots $& $\cdots $& F675W & 14.649(12) & 0.951(30)& 1.146(30)&  & \\
$\cdots $ & $\cdots $& $\cdots $& F555W & 14.741(21) & 0.746(40)& 0.996(40)&  & \\
$\cdots $ & $\cdots $& $\cdots $& F450W & 14.818(20) & 0.696(34)& 1.010(34)& $\cdots $ & 1.03$(^{+4}_{-5})$ \\
$\cdots $ & $\cdots $& $\cdots $& F336W & 14.950(13) & 0.515(94)& 0.920(94)& $\cdots $ & \\[+2mm]
\hline
\end{tabular} \\
\end{center}
\begin{tabular}{ll}
$^a$~Dots in the column show which filters were used for power law fitting. & $^d$~An upper limit. The aperture was placed on the estimated blob position in 2007.\\ 
$^b$~The row shows which pivot frequency was used for the normalization. & $^e$~The aperture was placed on the estimated blob position in 2005.\\
$^c$~The power law was fitted without the F336W data. & $^f$~The aperture was placed on the estimated blob position in 1999.\\ 
\end{tabular}
\end{table*}

%% file: table-Xray.tex
\begin{table}
\begin{center}
\caption{Results of X-ray spectral  fits for three regions of the 0540 
pulsar+PWN system. Photoelectric absorption along the line of sight towards 0540 was accounted
for using $N_H^{MW}=0.06\times 10^{22}$ (cm$^{-2}$), 
$N_H^{LMC}=0.5\times 10^{22}$ (cm$^{-2}$), 
$N_H^{SNR}=2.42\times 10^{19}$ (cm$^{-2}$). 
Note, that the spectra for the ``blob'' and ``anti-blob'' were extracted using the same 
apertures as their optical counterparts. See text for more details.} 
\label{xray:fits}
\begin{tabular}{lcrc}
\hline 
\hline
 Region & \multicolumn{2}{c}{Absorbed power law model} &  \\ 
\cline{2-3} 
 &  & & {$\chi^2$} \\
 &  $\alpha_{\nu}$ & normalization & per dof  \\
 &  &  {10$^{-4}$ ph cm$^{-2}$} &  \\
 &  & {s$^{-1}$ keV$^{-1}$} & \\
\hline
pulsar & $0.74\pm0.01$ & $22.97\pm2.53$ & 1.01   \\
``blob'' & $0.65\pm0.03$ & $2.94\pm0.05$  & 1.01   \\
``anti-blob'' & $0.73\pm0.04$ & $1.84\pm0.04$ & 0.93   \\
\hline 
\hline
\end{tabular} \\
\end{center}
\end{table}

%% file: Lundqvist_pwn0540_corrected.bbl
\begin{thebibliography}{}

 \bibitem[\protect\citeauthoryear{Anders \& Grevesse}{1989}]{Anders89} 
   Anders, E., Grevesse, N. 1989, Geochim. Cosmochim. Acta, 53, 197
   
 \bibitem[\protect\citeauthoryear{Benjamini \& Hochberg}{1995}]{Benjamini95} 
   Benjamini, Y., Hochberg, Y. 1995, J. R. Stat. Soc. B, 57, 289
  
 \bibitem[\protect\citeauthoryear{Bijaoui \& Ru\'e}{1995}]{Bijaoui95}
Bijaoui, A., Ru\'e, F. 1995, Signal Proc. 46, 229 

 \bibitem[\protect\citeauthoryear{Biretta \& McMaster}{1997}]{Biretta97} 
Biretta, J., McMaster, M. 1997, Instrument Science Report WFPC2 97-11

 \bibitem[\protect\citeauthoryear{Caraveo \etal}{2000}]{Car00} 
Caraveo, P. A., Mignani, R. P., De Luca, A., et al. 2000, in A decade of HST science,
eds. Mario Livio et al. (Baltimore) 105, [astro-ph/0009035] 

 \bibitem[\protect\citeauthoryear{Chanan \& Helfand}{1990}]{Chanan90} 
Chanan, G., A., Helfand, D., J. 1990, ApJ, 352, 167

 \bibitem[\protect\citeauthoryear{Chevalier}{2006}]{Chevalier06} 
 Chevalier, R. A. 2006, astro-ph/0607422

 \bibitem[\protect\citeauthoryear{De Luca \etal}{2007}]{DeLuca07} 
De Luca, A., Mignani, R. P., Caraveo, P. A., Bignami, G. F. 2007, ApJ, 667, 77

 \bibitem[\protect\citeauthoryear{Dickel \etal}{2002}]{Dickel02} 
Dickel, J., R., Mulligan, M., C., Klinger, R., J. et al. 2002, Neutron Stars in 
Supernova Remnants, ASP Conf. Series, 271, 195, Eds. Slane, P., O., Gaensler, B.

 \bibitem[\protect\citeauthoryear{Gotthelf \& Wang}{2000}]{GW00} 
Gotthelf, E. V., \& Wang, Q. D. 2000, ApJ, 532, L117

 \bibitem[\protect\citeauthoryear{Gr\"oningsson \etal}{2008}]{Gron08} 
Gr\"oningsson, P., Fransson, C., Lundqvist, P. et al. 2008, A\&A, 479, 761

 \bibitem[\protect\citeauthoryear{Hester \etal}{2002}]{Hester02} 
Hester, J. J., Mori, K.,, Burrows, D. et al. 2002, ApJ, 577, 49

 \bibitem[\protect\citeauthoryear{Hester}{2008}]{Hester08} 
Hester, J. J. 2008, ARA\&A, 46, 127 

 \bibitem[\protect\citeauthoryear{Jeffery}{1991}]{Jeffery91} 
 Jeffery, D. J. 1991, ApJS, 77, 405 

 \bibitem[\protect\citeauthoryear{Kaplan \etal}{2008}]{Kaplan08} 
Kaplan, D. L., Chatterjee, S., Gaensler, B. M., Anderson, J. 2008, ApJ, 677, 1201

 \bibitem[\protect\citeauthoryear{Kirshner \etal}{1989}]{Kirshner89} 
Kirshner, R. P., Morse, J. A., Winkler, P. F., Blair, W. P. 1989, ApJ, 342, 260

 \bibitem[\protect\citeauthoryear{Lundqvist \& Fransson}{19996}]{LF96} 
Lundqvist, P., Fransson, C. 1996, ApJ, 464, 942

 \bibitem[\protect\citeauthoryear{Mattila et al.}{2010}]{Matt10} 
Mattila, S., Lundqvist, P., Gr\"oningsson, P., Meikle, P. et al. 2010, sibmitted to ApJ, 
arXiv:1002.4195

 \bibitem[\protect\citeauthoryear{Melatos \etal}{2005}]{Melatos05} 
Melatos, A., Scheltus, D., Whiting, M. T., Eikenberry, S. S. et al. 2005, ApJ, 633, 931 

 \bibitem[\protect\citeauthoryear{Middleditch \etal}{1987}]{Middleditch87} 
Middleditch, R., N., Pennypacker, C., R., Burns, M., S. 1987, ApJ, 315, 142

 \bibitem[\protect\citeauthoryear{Mignani \etal}{2010}]{Mignani10}
Mignani, R. P., Sartori, A., De Luca, A., et al. 2010, astro-ph/1003.0786 

 \bibitem[\protect\citeauthoryear{Mori \etal}{2004}]{Mori04} 
Mori, K., Burrows, D. N., Hester, J. J. 2004, ApJ, 609, 186

 \bibitem[\protect\citeauthoryear{Morse \etal}{2006}]{Morse06} 
Morse, J. A., Smith, N., Blair, W. P., Kirshner, R. P. et al. 2006, ApJ, 644, 188

 \bibitem[\protect\citeauthoryear{Nomoto \& Hashimoto}{1988}]{Nom88} 
Nomoto, K., Hashimoto, M. 1988, Phys. Rep., 163, 13

 \bibitem[\protect\citeauthoryear{Panagia}{2005}]{Panagia05} 
Panagia, N. 2005  in the Proceedings of IAU 
Colloquium 192: Supernovae (10 years of SN1993J), eds. J.M. Marcaide \& K.W. 
Weiler (Springer Verlag), v.99, p. 585  [astro-ph/0309416]

 \bibitem[\protect\citeauthoryear{Petre \etal}{2007}]{Petre07} 
Petre, R., Hwang, U., Holt, S. S., et al. 2007, ApJ, 662, 997

 \bibitem[\protect\citeauthoryear{Riess}{2000}]{riess00} 
Riess, A., 2000, Instrument Science Report WFPC2, 04

 \bibitem[\protect\citeauthoryear{Saio \etal}{1988}]{Sai88} 
Saio, H., Nomoto, K., Kato, M. 1988, Nature, 334, 508

 \bibitem[\protect\citeauthoryear{Serafimovich \etal}{2004}]{Seraf04} 
 Serafimovich, N. I., Shibanov, Yu. A., Lundqvist, P., Sollerman, J. 2004, \aap, 425, 1041

 \bibitem[\protect\citeauthoryear{Serafimovich \etal}{2005}]{Seraf05} 
Serafimovich N. I., Lundqvist, P., Shibanov, Yu. A., Sollerman, J. 2005, AdSpR, 35, 1106

 \bibitem[\protect\citeauthoryear{Shigeyama \& Nomoto}{1990}]{Shig90} 
Shigeyama, T., Nomoto, K. 1990, ApJ, 360, 242

 \bibitem[\protect\citeauthoryear{Starck \& Murtagh}{1998}]{Starck98} 
 Starck, J.-L. \& Murtagh, F. 1998, Astronomical Society of the Pacific, 110, 744, 193-199
   
 \bibitem[\protect\citeauthoryear{Starck \& Murtagh}{2006}]{Starck06} 
Starck, J.-L. \& Murtagh, F. 2006, \aap library, ISBN-10 3-540-33024-0 2nd Edition Springer

 \bibitem[\protect\citeauthoryear{Tziamtzis \etal}{2009}]{Tziamtzis09} 
Tziamtzis, A., Schirmer, M., Lundqvist, P., Sollerman, J. 2009, \aap, 497, 167

 \bibitem[\protect\citeauthoryear{Wagner \& Seifert}{2000}]{Wagner00} 
Wagner, S., J., Seifert, W. 2000, Pulsar astronomy 2000 and beyond ASP Conf. ser., 
Kramer, M., Wex, N., Wielebinski, R. $-$ eds., 202, 315

 \bibitem[\protect\citeauthoryear{Weisskopf \etal}{2000}]{Weisskopf00} 
Weisskopf, M. C., Hester, J. J., Tennant, A. F. 2000, ApJ, 536L, 81
   
 \bibitem[\protect\citeauthoryear{Williams \etal}{2008}]{Williams08} 
Williams, B. J., Borkowski, K. J., Reynolds, S. P., et al. 2008, ApJ, 687, 1054

 \bibitem[\protect\citeauthoryear{Williams}{2010}]{Williams10} 
Williams, B. J. 2010, PhD Thesis (North Carolina State University, Raleigh)
[arXiv:1005.1296]

 \bibitem[\protect\citeauthoryear{Wilms}{2000}]{Wilms00} 
Wilms, J., Allen, A., McCray, R. 2000, ApJ, 542, 914
\end{thebibliography}
